\pdfoutput=1
\documentclass[zpreprint,zbstjhep,british]{zeus_K2.9paper}
\usepackage{graphicx}
\usepackage{epstopdf}
\usepackage{pdflscape}
\usepackage{rotating}
\usepackage{hepunits}
\begin{document}

\prepnum{DESY--22--107}
\prepdate{June 2022}

\zeustitle{%
  Measurement of the cross-section ratio {\boldmath $\sigma_{\psi(2S )}/\sigma_{J/\psi(1S )}$}
in exclusive photoproduction at HERA
}

\zeusauthor{ZEUS Collaboration\\
}

\zeusdate{}

\maketitle

\begin{abstract}\noindent
The exclusive photoproduction reactions $\gamma p \to J/\psi(1S) p$ and $\gamma p \to \psi(2S) p$ have been measured  
at an $ep$ centre-of-mass energy of 318\,GeV with the ZEUS detector at HERA using an integrated luminosity of 
373\,pb$^{-1}$.  
The measurement was made in the kinematic range $30 < W < 180$\,GeV, $Q^2 < 1$\,GeV$^2$ and 
$|t| < 1$\,GeV$^2$, where $W$ is the photon--proton centre-of-mass energy, $Q^2$ is the photon virtuality and $t$ 
is the squared four-momentum transfer at the proton vertex.  The decay channels used were $J/\psi(1S) \to \mu^+ \mu^-$,   
$\psi(2S) \to \mu^+ \mu^-$ and $\psi(2S) \to J/\psi(1S) \pi^+ \pi^-$ with subsequent decay $J/\psi(1S) \to \mu^+ \mu^-$.  The ratio of 
the production cross sections, $R = \sigma_{\psi(2S)} / \sigma_{J/\psi(1S)}$, has been measured as a function of 
$W$ and $|t|$ and compared to previous data in photoproduction and deep inelastic scattering and with predictions of 
QCD-inspired models of exclusive vector-meson production, which are in reasonable agreement with the data.
\end{abstract}

  \thispagestyle{empty}
\clearpage
%
%
%
%


\topmargin-1.cm
\evensidemargin-0.3cm
\oddsidemargin-0.3cm
\textwidth 16.cm
\textheight 680pt
\parindent0.cm
\parskip0.3cm plus0.05cm minus0.05cm
\def\3{\ss}
\newcommand{\address}{ }
\pagenumbering{Roman}
                                                   %
\begin{center}
{                      \Large  The ZEUS Collaboration              }
\end{center}

{\small\raggedright


I.~Abt$^{1}$, 
M.~Adamus$^{31}$, 
R. Aggarwal$^{2}$, 
V.~Aushev$^{3}$, 
O.~Behnke$^{4}$, 
A.~Bertolin$^{5}$, 
I.~Bloch$^{6}$, 
I.~Brock$^{7}$, 
N.H.~Brook$^{8, a}$, 
R.~Brugnera$^{9}$, 
A.~Bruni$^{10}$, 
P.J.~Bussey$^{11}$, 
A.~Caldwell$^{1}$, 
C.D.~Catterall$^{12}$, 
J.~Chwastowski$^{13}$, 
J.~Ciborowski$^{14, b}$, 
R.~Ciesielski$^{4, c}$, 
A.M.~Cooper-Sarkar$^{15}$, 
M.~Corradi$^{10, d}$, 
R.K.~Dementiev$^{16}$, 
S.~Dusini$^{5}$, 
J.~Ferrando$^{4}$, 
B.~Foster$^{15, e}$, 
E.~Gallo$^{17, f}$, 
D.~Gangadharan$^{18, g}$, 
A.~Garfagnini$^{9}$, 
A.~Geiser$^{4}$, 
G.~Grzelak$^{14}$, 
C.~Gwenlan$^{15}$, 
D.~Hochman$^{19}$, 
N.Z.~Jomhari$^{4}$, 
I.~Kadenko$^{3}$, 
U.~Karshon$^{19}$, 
P.~Kaur$^{20}$, 
R.~Klanner$^{17}$, 
U.~Klein$^{4, h}$, 
I.A.~Korzhavina$^{16}$, 
N.~Kovalchuk$^{17}$, 
M.~Kuze$^{21}$, 
B.B.~Levchenko$^{16}$, 
A.~Levy$^{22}$, 
B.~L\"ohr$^{4}$, 
E.~Lohrmann$^{17}$, 
A.~Longhin$^{9}$, 
F.~Lorkowski$^{4}$,
I.~Makarenko$^{4}$, 
J.~Malka$^{4, i}$, 
S.~Masciocchi$^{23, j}$, 
K.~Nagano$^{24}$, 
J.D.~Nam$^{25}$, 
Yu.~Onishchuk$^{3}$, 
E.~Paul$^{7}$, 
I.~Pidhurskyi$^{26}$, 
A.~Polini$^{10}$, 
M.~Przybycie\'n$^{27}$, 
A.~Quintero$^{25}$, 
I.~Rubinsky$^{4}$, 
M.~Ruspa$^{28}$, 
U.~Schneekloth$^{4}$, 
T.~Sch\"orner-Sadenius$^{4}$, 
I.~Selyuzhenkov$^{23}$, 
M.~Shchedrolosiev$^{4}$, 
L.M.~Shcheglova$^{16}$, 
I.O.~Skillicorn$^{11}$, 
W.~S{\l}omi\'nski$^{29}$, 
A.~Solano$^{30}$, 
L.~Stanco$^{5}$, 
N.~Stefaniuk$^{4}$, 
B.~Surrow$^{25}$, 
K.~Tokushuku$^{24}$, 
J.~Tomaszewska$^{14, m}$, 
A.~Trofymov$^{3,4}$, 
O.~Turkot$^{4, i}$, 
T.~Tymieniecka$^{31}$, 
A.~Verbytskyi$^{1}$, 
W.A.T.~Wan Abdullah$^{32}$, 
K.~Wichmann$^{4}$, 
M.~Wing$^{8, k}$, 
S.~Yamada$^{24}$, 
Y.~Yamazaki$^{33}$, 
A.F.~\.Zarnecki$^{14}$, 
O.~Zenaiev$^{4, l}$ 
\newpage


{\setlength{\parskip}{0.4em}
\makebox[3ex]{$^{1}$}
\begin{minipage}[t]{14cm}
{\it Max-Planck-Institut f\"ur Physik, M\"unchen, Germany}

\end{minipage}

\makebox[3ex]{$^{2}$}
\begin{minipage}[t]{14cm}
{\it DST-Inspire Faculty, Department of Technology, SPPU, India}

\end{minipage}

\makebox[3ex]{$^{3}$}
\begin{minipage}[t]{14cm}
{\it Department of Nuclear Physics, National Taras Shevchenko University of Kyiv, Kyiv, Ukraine}

\end{minipage}

\makebox[3ex]{$^{4}$}
\begin{minipage}[t]{14cm}
{\it Deutsches Elektronen-Synchrotron DESY, Notkestr.\ 85, 22607 Hamburg, Germany}

\end{minipage}

\makebox[3ex]{$^{5}$}
\begin{minipage}[t]{14cm}
{\it INFN Padova, Padova, Italy}~$^{A}$

\end{minipage}

\makebox[3ex]{$^{6}$}
\begin{minipage}[t]{14cm}
{\it Deutsches Elektronen-Synchrotron DESY, Platanenallee 6, 15738 Zeuthen, Germany}

\end{minipage}

\makebox[3ex]{$^{7}$}
\begin{minipage}[t]{14cm}
{\it Physikalisches Institut der Universit\"at Bonn,
Bonn, Germany}~$^{B}$

\end{minipage}

\makebox[3ex]{$^{8}$}
\begin{minipage}[t]{14cm}
{\it Physics and Astronomy Department, University College London,
London, United Kingdom}~$^{C}$

\end{minipage}

\makebox[3ex]{$^{9}$}
\begin{minipage}[t]{14cm}
{\it Dipartimento di Fisica e Astronomia dell' Universit\`a and INFN,
Padova, Italy}~$^{A}$

\end{minipage}

\makebox[3ex]{$^{10}$}
\begin{minipage}[t]{14cm}
{\it INFN Bologna, Bologna, Italy}~$^{A}$

\end{minipage}

\makebox[3ex]{$^{11}$}
\begin{minipage}[t]{14cm}
{\it School of Physics and Astronomy, University of Glasgow,
Glasgow, United Kingdom}~$^{C}$

\end{minipage}

\makebox[3ex]{$^{12}$}
\begin{minipage}[t]{14cm}
{\it Department of Physics, York University, Ontario, Canada M3J 1P3}~$^{D}$

\end{minipage}

\makebox[3ex]{$^{13}$}
\begin{minipage}[t]{14cm}
{\it The Henryk Niewodniczanski Institute of Nuclear Physics, Polish Academy of \\
Sciences, Krakow, Poland}

\end{minipage}

\makebox[3ex]{$^{14}$}
\begin{minipage}[t]{14cm}
{\it Faculty of Physics, University of Warsaw, Warsaw, Poland}

\end{minipage}

\makebox[3ex]{$^{15}$}
\begin{minipage}[t]{14cm}
{\it Department of Physics, University of Oxford,
Oxford, United Kingdom}~$^{C}$

\end{minipage}

\makebox[3ex]{$^{16}$}
\begin{minipage}[t]{14cm}
{\it Affiliated with an institute covered by a current or former 
collaboration agreement with DESY}

\end{minipage}

\makebox[3ex]{$^{17}$}
\begin{minipage}[t]{14cm}
{\it Hamburg University, Institute of Experimental Physics, Hamburg,
Germany}~$^{E}$

\end{minipage}

\makebox[3ex]{$^{18}$}
\begin{minipage}[t]{14cm}
{\it Physikalisches Institut of the University of Heidelberg, Heidelberg, Germany}

\end{minipage}

\makebox[3ex]{$^{19}$}
\begin{minipage}[t]{14cm}
{\it Department of Particle Physics and Astrophysics, Weizmann
Institute, Rehovot, Israel}

\end{minipage}

\makebox[3ex]{$^{20}$}
\begin{minipage}[t]{14cm}
{\it Sant Longowal Institute of Engineering and Technology, Longowal, Punjab, India}

\end{minipage}

\makebox[3ex]{$^{21}$}
\begin{minipage}[t]{14cm}
{\it Department of Physics, Tokyo Institute of Technology,
Tokyo, Japan}~$^{F}$

\end{minipage}

\makebox[3ex]{$^{22}$}
\begin{minipage}[t]{14cm}
{\it Raymond and Beverly Sackler Faculty of Exact Sciences, School of Physics, \\
Tel Aviv University, Tel Aviv, Israel}~$^{G}$

\end{minipage}

\makebox[3ex]{$^{23}$}
\begin{minipage}[t]{14cm}
{\it GSI Helmholtzzentrum f\"{u}r Schwerionenforschung GmbH, Darmstadt, Germany}

\end{minipage}

\makebox[3ex]{$^{24}$}
\begin{minipage}[t]{14cm}
{\it Institute of Particle and Nuclear Studies, KEK,
Tsukuba, Japan}~$^{F}$

\end{minipage}

\makebox[3ex]{$^{25}$}
\begin{minipage}[t]{14cm}
{\it Department of Physics, Temple University, Philadelphia, PA 19122, USA}~$^{H}$

\end{minipage}

\makebox[3ex]{$^{26}$}
\begin{minipage}[t]{14cm}
{\it Institut f\"ur Kernphysik, Goethe Universit\"at, Frankfurt am Main, Germany}

\end{minipage}

\makebox[3ex]{$^{27}$}
\begin{minipage}[t]{14cm}
{\it AGH University of Science and Technology, Faculty of Physics and Applied Computer
Science, Krakow, Poland}

\end{minipage}

\makebox[3ex]{$^{28}$}
\begin{minipage}[t]{14cm}
{\it Universit\`a del Piemonte Orientale, Novara, and INFN, Torino,
Italy}~$^{A}$

\end{minipage}

\makebox[3ex]{$^{29}$}
\begin{minipage}[t]{14cm}
{\it Department of Physics, Jagellonian University, Krakow, Poland}~$^{I}$

\end{minipage}

\makebox[3ex]{$^{30}$}
\begin{minipage}[t]{14cm}
{\it Universit\`a di Torino and INFN, Torino, Italy}~$^{A}$

\end{minipage}

\makebox[3ex]{$^{31}$}
\begin{minipage}[t]{14cm}
{\it National Centre for Nuclear Research, Warsaw, Poland}

\end{minipage}

\makebox[3ex]{$^{32}$}
\begin{minipage}[t]{14cm}
{\it National Centre for Particle Physics, Universiti Malaya, 50603 Kuala Lumpur, Malaysia}~$^{J}$

\end{minipage}

\makebox[3ex]{$^{33}$}
\begin{minipage}[t]{14cm}
{\it Department of Physics, Kobe University, Kobe, Japan}~$^{F}$

\end{minipage}

}

\vspace{3em}


{\setlength{\parskip}{0.4em}\raggedright
\makebox[3ex]{$^{ A}$}
\begin{minipage}[t]{14cm}
 supported by the Italian National Institute for Nuclear Physics (INFN) \
\end{minipage}

\makebox[3ex]{$^{ B}$}
\begin{minipage}[t]{14cm}
 supported by the German Federal Ministry for Education and Research (BMBF), under
 contract No.\ 05 H09PDF\
\end{minipage}

\makebox[3ex]{$^{ C}$}
\begin{minipage}[t]{14cm}
 supported by the Science and Technology Facilities Council, UK\
\end{minipage}

\makebox[3ex]{$^{ D}$}
\begin{minipage}[t]{14cm}
 supported by the Natural Sciences and Engineering Research Council of Canada (NSERC) \
\end{minipage}

\makebox[3ex]{$^{ E}$}
\begin{minipage}[t]{14cm}
 supported by the German Federal Ministry for Education and Research (BMBF), under
 contract No.\ 05h09GUF, and the SFB 676 of the Deutsche Forschungsgemeinschaft (DFG) \
\end{minipage}

\makebox[3ex]{$^{ F}$}
\begin{minipage}[t]{14cm}
 supported by the Japanese Ministry of Education, Culture, Sports, Science and Technology
 (MEXT) and its grants for Scientific Research\
\end{minipage}

\makebox[3ex]{$^{ G}$}
\begin{minipage}[t]{14cm}
 supported by the Israel Science Foundation\
\end{minipage}

\makebox[3ex]{$^{ H}$}
\begin{minipage}[t]{14cm}
 supported in part by the Office of Nuclear Physics within the U.S.\ DOE Office of Science
\end{minipage}

\makebox[3ex]{$^{ I}$}
\begin{minipage}[t]{14cm}
supported by the Polish National Science Centre (NCN) grant no.\ DEC-2014/13/B/ST2/02486
\end{minipage}

\makebox[3ex]{$^{ J}$}
\begin{minipage}[t]{14cm}
 supported by HIR grant UM.C/625/1/HIR/149 and UMRG grants RU006-2013, RP012A-13AFR and RP012B-13AFR from
 Universiti Malaya, and ERGS grant ER004-2012A from the Ministry of Education, Malaysia\
\end{minipage}

}

\pagebreak[4]
{\setlength{\parskip}{0.4em}


\makebox[3ex]{$^{ a}$}
\begin{minipage}[t]{14cm}
now at University of Bath, United Kingdom\
\end{minipage}

\makebox[3ex]{$^{ b}$}
\begin{minipage}[t]{14cm}
also at Lodz University, Poland\
\end{minipage}

\makebox[3ex]{$^{ c}$}
\begin{minipage}[t]{14cm}
now at Rockefeller University, New York, NY 10065, USA\
\end{minipage}

\makebox[3ex]{$^{ d}$}
\begin{minipage}[t]{14cm}
now at INFN Roma, Italy\
\end{minipage}

\makebox[3ex]{$^{ e}$}
\begin{minipage}[t]{14cm}
also at DESY and University of Hamburg, Hamburg, Germany and supported by a Leverhulme Trust Emeritus Fellowship\
\end{minipage}

\makebox[3ex]{$^{ f}$}
\begin{minipage}[t]{14cm}
also at DESY, Hamburg, Germany\
\end{minipage}

\makebox[3ex]{$^{ g}$}
\begin{minipage}[t]{14cm}
now at now at University of Houston, Houston, TX 77004, USA\
\end{minipage}

\makebox[3ex]{$^{ h}$}
\begin{minipage}[t]{14cm}
now at University of Liverpool, United Kingdom\
\end{minipage}

\makebox[3ex]{$^{ i}$}
\begin{minipage}[t]{14cm}
now at European X-ray Free-Electron Laser facility GmbH, Hamburg, Germany\
\end{minipage}

\makebox[3ex]{$^{ j}$}
\begin{minipage}[t]{14cm}
also at Physikalisches Institut of the University of Heidelberg, Heidelberg,  Germany\
\end{minipage}

\makebox[3ex]{$^{ k}$}
\begin{minipage}[t]{14cm}
also supported by DESY, Hamburg, Germany\
\end{minipage}

\makebox[3ex]{$^{ l}$}
\begin{minipage}[t]{14cm}
now at CERN, Geneva, Switzerland\
\end{minipage}

\makebox[3ex]{$^{ m}$}
\begin{minipage}[t]{14cm}
now at Polish Air Force Academy in Deblin\
\end{minipage}


}

}


\cleardoublepage
\pagenumbering{arabic}
%
%
\pagestyle{scrheadings}
 
\section{Introduction}
\label{sec-int}

\begin{figure}[p]
\vfill
\begin{center}
\includegraphics[width=12.0cm]{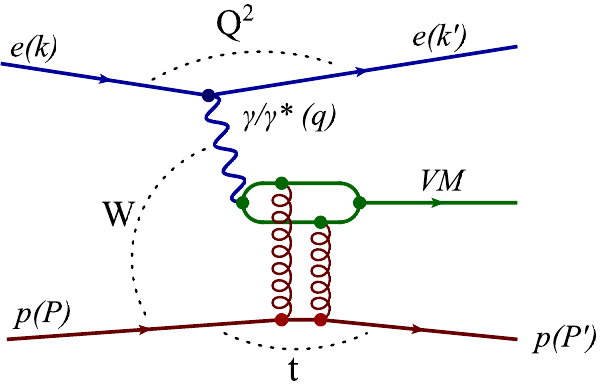}
\put(-320,218){\makebox(0,0)[tl]{(a)}}\\
\vspace{2cm}
\includegraphics[width=12.0cm]{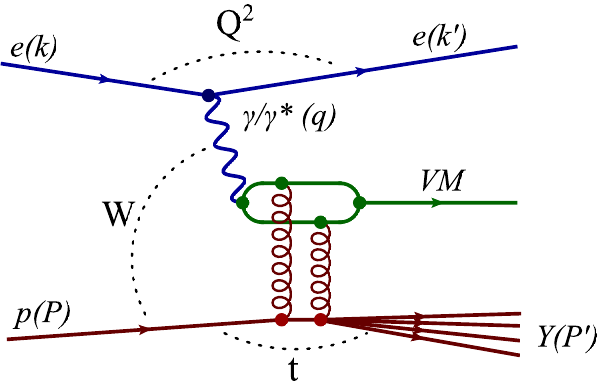}
\put(-320,218){\makebox(0,0)[tl]{(b)}}
\end{center}
 \caption{
Schematic representation of (a) exclusive and (b) proton-dissociative vector-meson production in $ep$ scattering.   For a description of 
the kinematic variables, see Section~\ref{sec-int}.
}
\label{fig-diagram}
\vfill
\end{figure}

The exclusive photoproduction of vector mesons leads to a simple final-state system, as illustrated in Fig.~\ref{fig-diagram}(a).  
The clean environment and the large masses of the $J/\psi(1S)$ and $\psi(2S)$ mesons facilitate measurements that   
provide insight into the dynamics of a hard process.  The $J/\psi (1S)$ and the $\psi (2S)$ have the same quark content but 
different radial distributions of the wave functions, and their mass difference is small.
Therefore, this measurement allows QCD-inspired predictions of the wave-function dependence of the respective 
cross sections to be tested.  A suppression of the $\psi (2S)$ cross section relative to the $J/\psi(1S)$ is expected, as the 
$\psi(2S)$ wave function has a radial node close to the typical transverse separation of the virtual $c \bar{c}$ pair.  The 
process is also sensitive to the gluon density in the proton.

The exclusive (also referred to as elastic) production of a vector meson in $ep$ collisions, in which the proton remains 
intact, is shown in 
Fig.~\ref{fig-diagram}(a).  The proton-dissociative process, where the proton breaks up into a hadronic state denoted as 
$Y$, is shown in Fig.~\ref{fig-diagram}(b).  The two processes have similar experimental signatures and so when the system $Y$ remains 
undetected, proton-dissociative events form a significant background.  The following kinematic variables are used to 
characterise these processes.  The negative squared four-momentum of the exchanged photon, $Q^2$, is equal to 
$-q^2 = -(k-k^\prime)^2$, where $k$ and $k^\prime$ are the four-momenta of the incoming and outgoing lepton.  As 
$Q^2 \approx 0$\,GeV$^2$ in photoproduction and the transverse momentum of the vector meson is small in the 
present measured kinematic region, the hard QCD scale is provided by the squared mass of the vector meson, 
$M_V^2$.  The photon--proton centre-of-mass energy, $W$, is given by $W^2 = (q+P)^2$, where $P$ is the 
four-momentum of the incoming proton.  The squared four-momentum transfer at the proton vertex, $t$, is given by 
$t = (P-P^\prime)^2$, where $P^\prime$ is the four-momentum of the outgoing proton (or dissociative state $Y$).

At the HERA $ep$ collider, the ZEUS collaboration has previously measured the exclusive production of $J/\psi(1S)$ and $\psi(2S)$ mesons in 
deep inelastic scattering (DIS)~\cite{np:b909:934}.  The H1 collaboration has also measured exclusive production of $J/\psi(1S)$ and $\psi(2S)$ mesons in 
DIS~\cite{epj:c10:373}, as well as in photoproduction~\cite{pl:b541:251,pl:b421:385}.  In these analyses, the ratio of the 
$\psi(2S)$ to $J/\psi(1S)$ production cross sections, where some of the systematic uncertainties are expected to cancel, 
was measured and compared with QCD-inspired models.  
These previous data exhibit an increase in the ratio with increasing $Q^2$ that is 
described by many of the QCD models.  No dependence of the ratio with $W$ and $|t|$ was observed, although the $|t|$ 
dependence has so far been measured only in DIS.

In this paper, a new measurement of the ratio of the photoproduction cross sections of the exclusive reactions 
$e p \rightarrow e \psi(2S ) p$ and $e p \rightarrow e J/\psi(1S ) p$ is presented.  The ratio is measured differentially 
as a function of $W$ and $|t|$.  The decay channels used were 
$J/\psi(1S) \to \mu^+ \mu^-$, $\psi(2S)\to \mu^+\mu^-$, and $\psi(2S)\,\to J/\psi(1S)\,\pi^+\pi^-$ with the subsequent 
decay $J/\psi(1S )\to \mu^+\mu^-$.

 \section{Experimental set-up}
  \label{sec-exp}

The measurement is based on data collected with the ZEUS detector at the HERA collider during the period 2003--2007, 
corresponding to an integrated luminosity of 373\,pb$^{-1}$.  During this period, the HERA accelerator collided an 
electron\footnote{Hereafter, ``electron'' refers to both electrons and positrons unless otherwise stated.} beam of energy 
27.5\,GeV with a proton beam of 920\,GeV, yielding an $ep$ centre-of-mass energy of 318\,GeV. 

 A detailed description of the ZEUS detector can be found elsewhere~\cite{zeus:1993:bluebook}. A brief outline of the 
 components that are most relevant for this analysis is given below.

In the kinematic range of the analysis, charged particles were mainly tracked in the central tracking detector 
(CTD)~\cite{nim:a279:290, npps:b32:181, nim:a338:254} and the microvertex detector (MVD)~\cite{nim:a581:656}.  
These components operated in a magnetic field of \unit{1.43}{\tesla} provided by a thin superconducting solenoid. The 
CTD consisted of 72~cylindrical drift-chamber layers, organised in nine superlayers covering the 
polar-angle\footnote{The ZEUS coordinate system is a right-handed Cartesian system, with the $Z$ axis pointing in the 
nominal proton beam direction, referred to as the ``forward direction'', and the $X$ axis pointing left towards the centre 
of HERA.  The coordinate origin is at the centre of the CTD.  The pseudorapidity is defined as 
$\eta=-\ln\left(\tan\frac{\theta}{2}\right)$, where the polar angle, $\theta$, is measured with respect to the $Z$ axis. The 
azimuthal angle, $\varphi$, is measured with respect to the $X$ axis.} region 
\mbox{$\unit{15}{\degree}<\theta<\unit{164}{\degree}$}.  The MVD silicon tracker consisted of a barrel (BMVD) and a 
forward (FMVD) section. The BMVD contained three layers and provided polar-angle coverage for tracks from 
$\unit{30}{\degree}$ to $\unit{150}{\degree}$. The four-layer FMVD extended the polar-angle coverage in the forward 
region to $\unit{7}{\degree}$. After alignment, the single-hit resolution of the MVD was \unit{24}{\micron}. The transverse 
distance of closest approach (DCA) of a track to the nominal vertex in the $X$--$Y$ plane was measured to have a resolution, averaged 
over the azimuthal angle, of \unit{$(46 \oplus 122 / p_T)$}{\micron}, with $p_T$ in \GeV~denoting the momentum 
transverse to the beam axis.  For CTD--MVD tracks that pass through all nine CTD superlayers, the momentum 
resolution was $\sigma(p_T)/p_T = 0.0029 p_T \oplus 0.0081 \oplus 0.0012/p_T$, with $p_T$ in \GeV.

The high-resolution uranium--scintillator calorimeter (CAL)~\cite{nim:a309:77, nim:a309:101, nim:a321:356, nim:a336:23} 
consisted of three parts: the forward (FCAL), the barrel (BCAL) and the rear (RCAL) calorimeters. Each part was 
subdivided transversely into towers and longitudinally into one electromagnetic section (EMC) and either one (in RCAL) 
or two (in BCAL and FCAL) hadronic sections (HAC). The smallest subdivision of the calorimeter was called a cell.  Adjacent cells 
were combined to form clusters.  The 
CAL energy resolutions, as measured under test-beam conditions, were $\sigma(E)/E=0.18/\sqrt{E}$ for electrons and 
$\sigma(E)/E=0.35/\sqrt{E}$ for hadrons, with $E$ in GeV.  

The muon system consisted of rear, barrel (R/BMUON) and forward (FMUON) tracking detectors. The R/BMUON consisted 
of limited-streamer (LS) tube chambers placed behind the RCAL (BCAL), inside and outside a magnetised iron yoke 
surrounding the CAL. The barrel and rear muon chambers covered polar angles from 34$^\circ$ to 135$^\circ$ and from 
135$^\circ$ to 171$^\circ$, respectively. The FMUON consisted of six trigger planes of LS tubes and four planes of drift 
chambers covering the angular region from 5$^\circ$ to 32$^\circ$. The muon system exploited the magnetic field of the 
iron yoke and, in the forward direction, of two iron toroids magnetised to about $1.6$\,T to provide an independent 
measurement of the muon momenta.

The iron yoke surrounding the CAL was instrumented with proportional drift chambers to form the backing calorimeter 
(BAC)~\cite{nim:a300:480}.  The BAC consisted of 5142 aluminium chambers inserted into the gaps between $7.3\,{\rm cm}$ 
thick iron plates (10, 9 and 7 layers in the forward, barrel and rear directions, respectively). The chambers were typically 
$5\,{\rm m}$ long and had a wire spacing of $1.5\,{\rm cm}$. The anode wires were covered by $50\,{\rm cm}$ long cathode 
pads. The BAC was equipped with energy readout and position-sensitive readout for muon tracking. The former was based 
on 1692 pad towers ($50 \times 50\,{\rm cm^2}$), providing an energy resolution of $\sigma(E)/E \approx 100\%/\sqrt E$, 
with $E$ in GeV. The position information from the wires allowed the reconstruction of muon trajectories in two dimensions 
($XY$ in the barrel and $YZ$ in the endcaps) with a spatial accuracy of a few mm.

The luminosity was measured using the Bethe--Heitler reaction $ep\,\rightarrow\, e\gamma p$ by a luminosity detector 
which consisted of independent lead--scintillator calorimeter~\cite{desy-92-066, zfp:c63:391, Andruszkow:2001jy} and magnetic 
spectrometer~\cite{nim:a565:572,Adamczyk:2013ewk} systems.


\section{Monte Carlo simulations}
\label{sec:mc}

Free parameters within the Monte Carlo (MC) simulations, which control the kinematic dependences of the reactions of interest, 
have been tuned to previous data.  
The values have been checked here and either used where appropriate or tuned to the data presented in this paper.

The {\sc Diffvm}~\cite{proc:mc:1998:396} MC program was used for simulating the photoproduction of exclusive 
heavy vector mesons, $ep \rightarrow e V p$, where $V$ denotes the produced vector meson.  For the event generation, 
the following cross-section parameterisations were used: 

\begin{itemize}

\item $(1 + Q^2/M_V^2)^{-1.5}$ for the dependence on $Q^2$; 

\item $W^\delta$, with $\delta = 0.67$~\cite{epj:c73:2466} and $\delta = 1.1$~\cite{levy:2009delta} for the dependence on $W$ of the total cross section for $J/\psi(1S)$ and $\psi (2S)$ events, respectively; 
         
\item $\exp(-b\,|t|)$, with $b = 4.6$\,GeV$^{-2}$ and $b = 4.3$\,GeV$^{-2}$ for the dependence on $|t|$ for $J/\psi(1S)$ and $\psi (2S)$ 
         events, respectively~\cite{pl:b541:251};  
         
\item the $b$ values were re-weighted according to the formula $b^\prime = b + 4 \, \alpha^\prime \ln(W/W_0)$, where 
         $W_0 = 90$\,GeV and $\alpha^\prime$ is determined to be 0.12\,GeV$^{-2}$~\cite{epj:c24:345}; 

\item $s$-channel helicity conservation for the production of $V \rightarrow \mu ^+ \mu ^-$; 

\item a reweighting for the pion phase space~\cite{pl:b61:183} using the function $(M(\pi^+,\pi^-)^2 - 4 \, M_\pi^2)^2$, where 
        $M(\pi^+,\pi^-)$ is the invariant mass of the two pions in the $\psi (2S) \to J/\psi (1S ) \pi ^+ \pi ^-$ decay and $M_\pi$ is the 
        mass of the charged pion.

\end{itemize}

Proton-dissociative $J/\psi (1S)$ and $\psi (2S)$ events were also simulated with the {\sc Diffvm} MC program with 
parameters:

\begin{itemize}

\item $\delta = 0.42$~\cite{epj:c73:2466} and $\delta = 0.70$ (tuned here) for the $W$ dependence for $J/\psi(1S)$ and $\psi (2S)$ events, 
respectively; 

\item $b = 1.0$\,GeV$^{-2}$ and $b = 0.7$\,GeV$^{-2}$ for the respective $|t|$ dependences~\cite{pl:b541:251}; 

\item the dependence on the mass of the dissociated proton system, $M_Y$, was simulated as $1/M_Y^\beta$, with $\beta = 2.4$ (tuned 
here) for both $J/\psi(1S)$ and $\psi(2S)$ production above the proton-resonance region, i.e.\ $M_Y \gtrsim 2$\,GeV.  

\end{itemize}

Non-resonant electroweak dimuon production (Bethe--Heitler process) was simulated using the program  
{\sc Grape}~\cite{cpc:136:126}.  The MC sample contains both exclusive and proton-dissociative events.

The generated MC events were passed through the ZEUS detector and trigger simulation programs based on 
{\sc Geant} 3~\cite{tech:cern-dd-ee-84-1}.  They were then reconstructed and analysed with the same programs as used for the 
data.


\section{Event selection and signal extraction }
\label{sec-event_selection}

\subsection{Event selection}

Events that contained signals from the decay products of the $\psi(2S)$ or $J/\psi(1S)$ but no other activity in the central ZEUS detector 
were selected.  Only final states containing muons were considered.

A three-level trigger system~\cite{zeus:1993:bluebook, nim:a580:1257, desy-92-150b} was used to select events online.  
The principal requirement for muon-candidate events was at least one CTD track matched to a cluster consistent with a minimum-ionising 
particle in the CAL and associated with a F/B/RMUON deposit or with a muon signal in the BAC.

To select offline events containing exclusively produced $J/\psi(1S)$ or $\psi(2S)$ vector mesons in photoproduction, the following 
additional requirements were imposed:

\begin{itemize}

\item the $Z$ coordinate of the event vertex reconstructed from the tracks was required to be within $\pm 30$\,cm of
the nominal $ep$ interaction point and the transverse distance of the event vertex from the nominal $ep$ interaction point was required 
to be within 0.03\,cm;

\item events with an identified electron with energy above 5\,GeV, as reconstructed using an algorithm based on a neural 
network~\cite{nim:a365:508}, were rejected.  This removed DIS events with $Q^2 > 1$\,GeV$^2$;

\item the sum of energy in the FCAL cells immediately surrounding the beam-pipe hole ($\theta<0.12$\,rad) had to be smaller 
than 1\,GeV to suppress contamination from proton-dissociative events;

\item the photon--proton centre-of-mass energy was required to be in the range $30 < W < 180$\,GeV, where $W$ is reconstructed 
from the initial proton-beam energy, $E_p$, and the difference in energy and $Z$ component of momentum, $p_Z$, of the vector-meson candidate, 
$V$, as $W = \sqrt{2 E_p (E-p_Z)_V}$;

\item the squared four-momentum transfer at the proton vertex was required to be in the range $|t| < 1$\,GeV$^2$, where $t$ is 
reconstructed from the transverse momentum of the vector meson, $p_{T,V}$, as $t = -(p_{T,V})^2$. This requirement significantly 
reduced the remaining fraction of proton-dissociative events;

\item each track considered was required to produce hits in the first CTD superlayer or in the MVD and cross at least three 
CTD superlayers.  
These requirements effectively limit the pseudorapidity range of each track to $-1.9 < \eta < 1.9$ and ensured the selection 
of tracks with good momentum resolution;

\item two oppositely-charged tracks, each with $p_T > 1$\,GeV, matched to the vertex were required in the event.  Each  
of these tracks was matched with a cluster in the CAL.  The cluster was required to be consistent with a muon identified with an algorithm based 
on a neural network~\cite{nim:a453:336}.  At least one of these tracks had to be associated with a F/B/RMUON signal or with a muon 
signal in the BAC found using the GMUON algorithm with muon quality $\ge 1$~\cite{thesis:bloch:2005};

\item for the selection of $J/\psi(1S)\,\to\,\mu^+\mu^-$ and $\psi(2S)\,\to\,\mu^+\mu^-$ events, no additional tracks were allowed.   
Cosmic-ray events were rejected by requiring $t_{\rm down} - t_{\rm up} < 8$\,ns, where $t_{\rm down}$ and $t_{\rm up}$ represent 
the calorimeter signal times in the lower and upper halves of the CAL.   Additionally, $\cos\alpha > -0.985$ was required, where $\alpha$ is the 
angle between the momentum vectors of the candidate $\mu^+$ and $\mu^-$;

\item for the selection of $\psi(2S)\,\to\,J/\psi(1S )\,\pi^+\pi^-$ events, exactly two additional oppositely-charged tracks were 
required.  Their momenta were required to be lower than those of the muons.  Each track was required to have a transverse 
momentum above 0.12\,GeV.  No explicit vertex association was required for these two tracks;

\item the energy of any additional CAL cluster not associated with a muon candidate, or with a pion candidate in the case of 
$\psi(2S)\,\to\,J/\psi(1S )\,\pi^+\pi^-$ events, was required to be less than 0.5\,GeV.  This ensured that events with 
other produced neutral particles were rejected while events with clusters in the CAL consistent with noise only were not rejected.

\end{itemize}

After rejection of DIS events, a study of generator-level events from the {\sc Diffvm} MC sample yielded a median $Q^2$ of 
about $3 \times 10^{-5}$\,GeV$^2$.  A similar study showed that 99\% of proton-dissociative events remaining after the above 
requirements had a diffractive mass $M_Y \lesssim 5$\,GeV.

\subsection{Signal extraction}
\label{sec:signal}

In the following, the signal extraction for the $\mu^+\mu^-$ and $\mu^+\mu^-\pi^+\pi^-$ final states are discussed separately.

\begin{figure}[p]
\vfill
\begin{center}
\includegraphics[trim={0cm 0cm 1cm 0.cm},clip,width=0.49\textwidth]{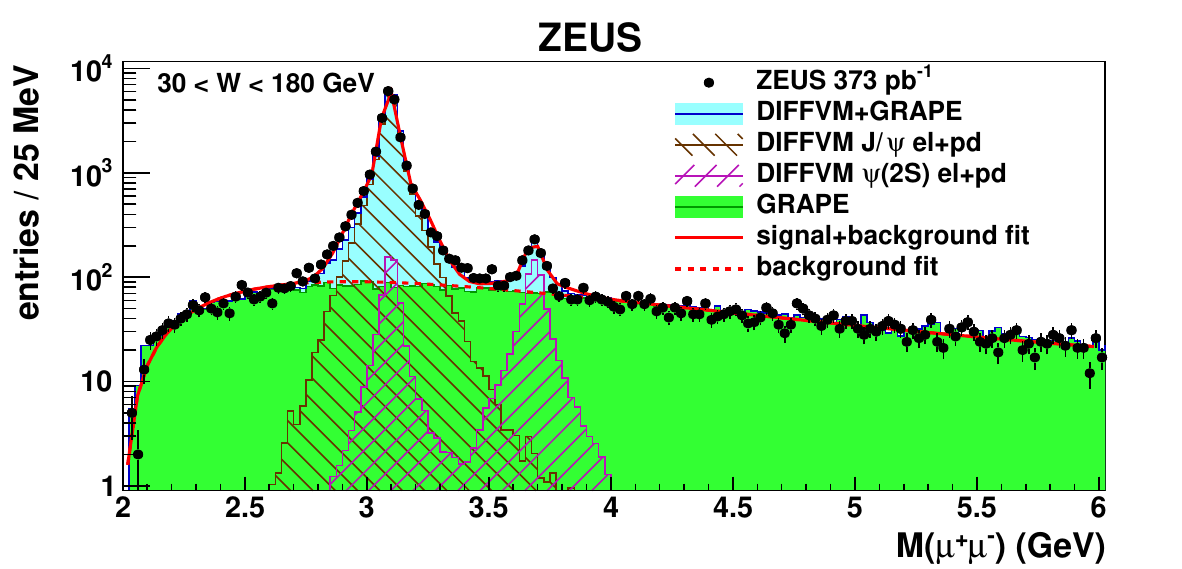}
\put(-185,90){\makebox(0,0)[tl]{(a)}}
\includegraphics[trim={0cm 0cm 1cm 0.cm},clip,width=0.49\textwidth]{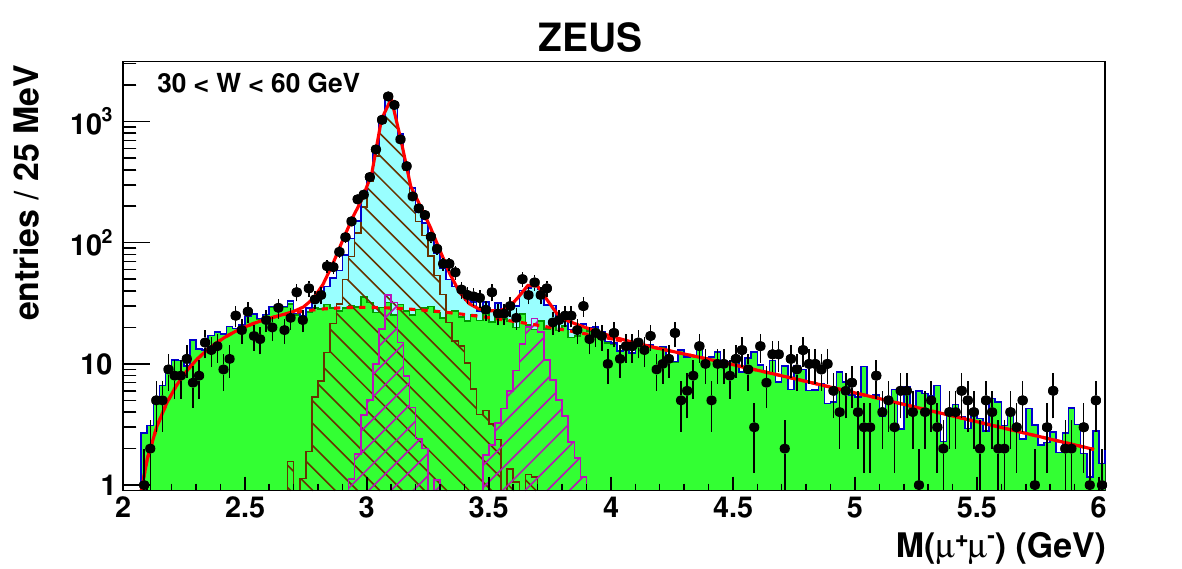}
\put(-185,90){\makebox(0,0)[tl]{(b)}}\\
\includegraphics[trim={0cm 0cm 1cm 0.cm},clip,width=0.49\textwidth]{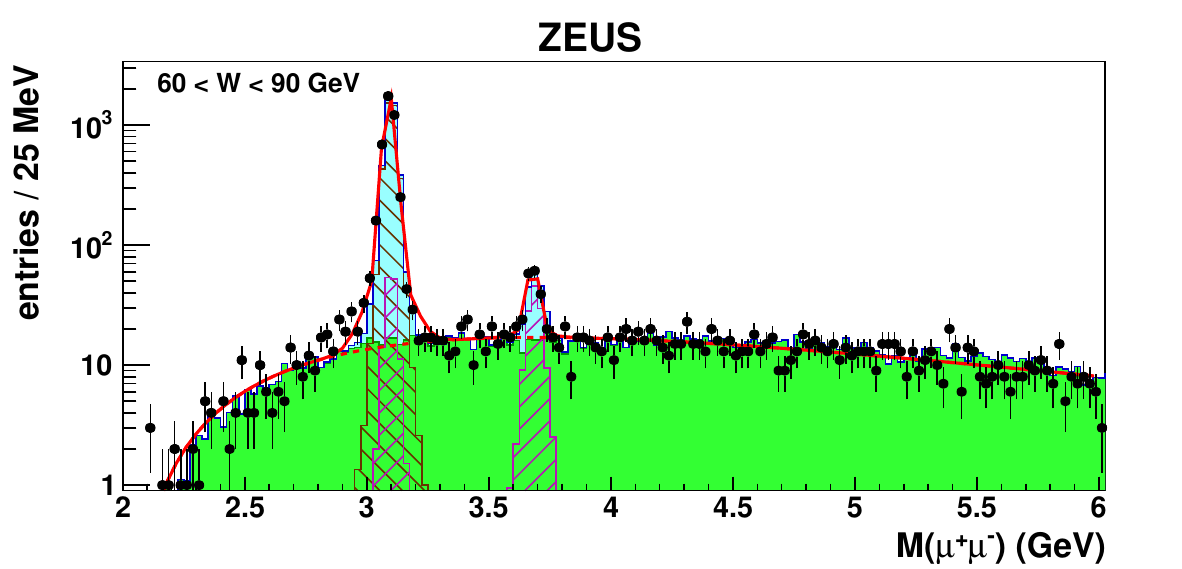}
\put(-185,90){\makebox(0,0)[tl]{(c)}}
\includegraphics[trim={0cm 0cm 1cm 0.cm},clip,width=0.49\textwidth]{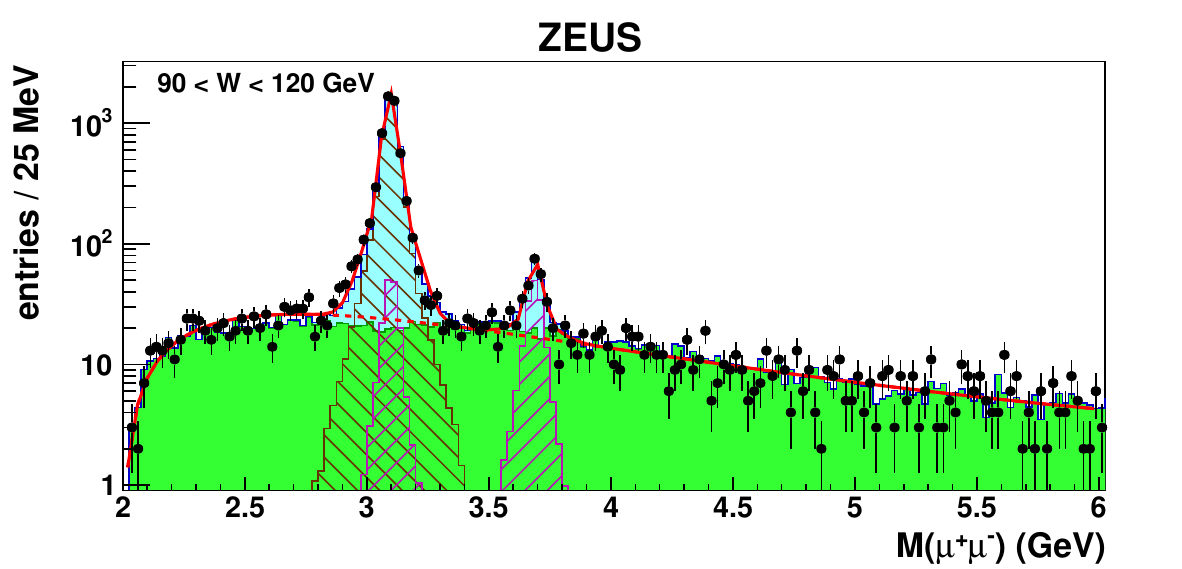}
\put(-185,90){\makebox(0,0)[tl]{(d)}}\\
\includegraphics[trim={0cm 0cm 1cm 0.cm},clip,width=0.49\textwidth]{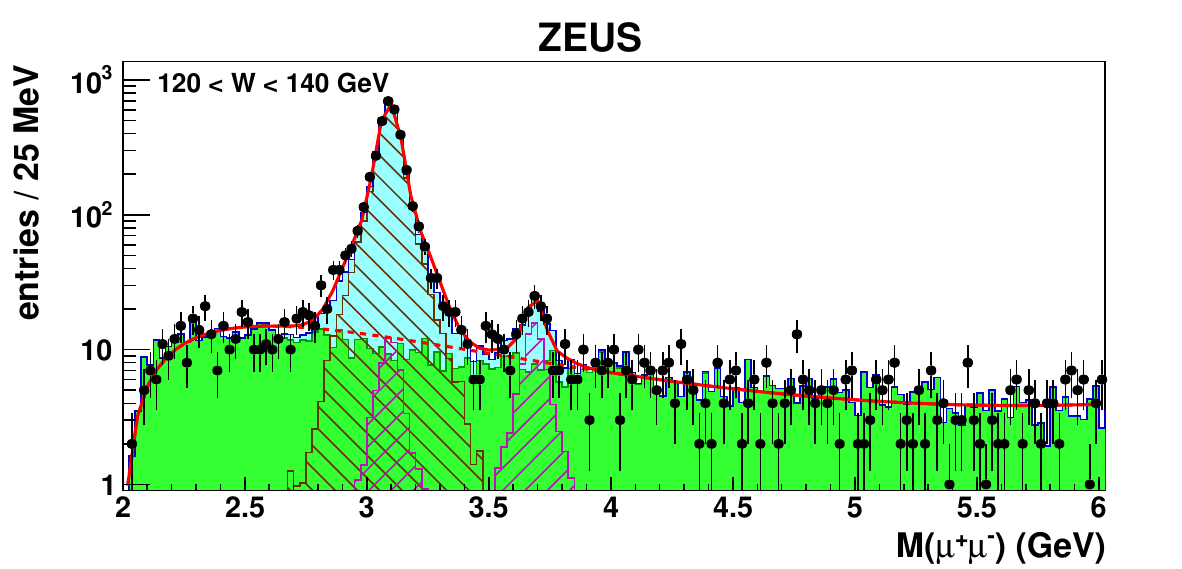}
\put(-185,90){\makebox(0,0)[tl]{(e)}}
\includegraphics[trim={0cm 0cm 1cm 0.cm},clip,width=0.49\textwidth]{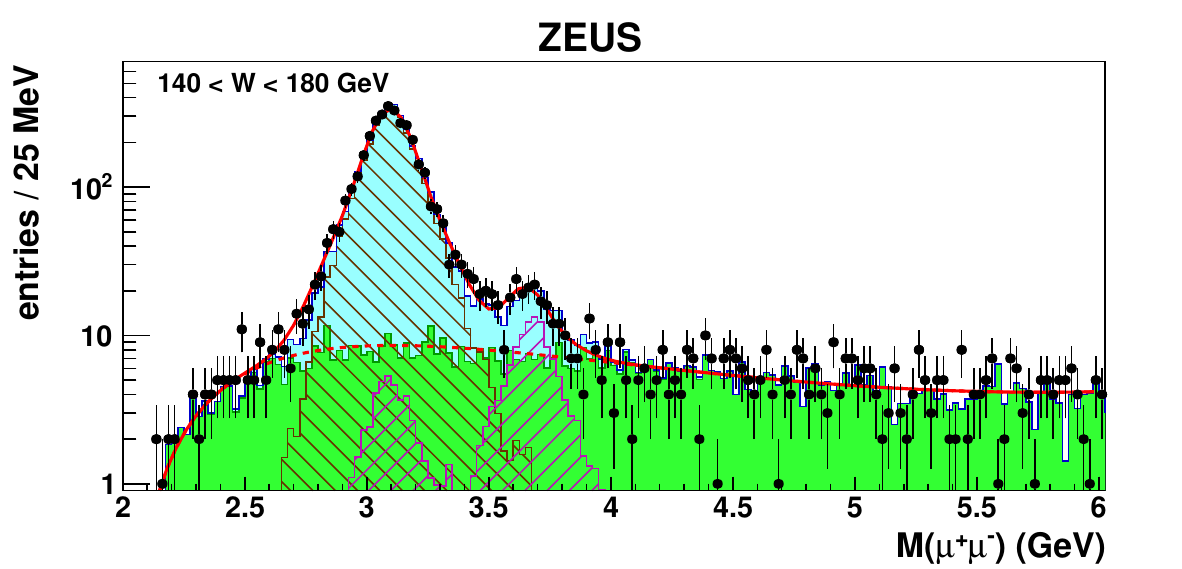}
\put(-185,90){\makebox(0,0)[tl]{(f)}}
\end{center}
\vspace{-0.5cm}
 \caption{
Measured invariant mass distribution, $M(\mu^+\mu^-)$, of dimuon pairs (solid circles) in selected photoproduction events with error bars denoting statistical uncertainties.  
The data are shown in (a) the full $W$ range, $30 < W < 180$\,GeV, and (b) -- (f) finer $W$ intervals within the full range. 
Monte Carlo distributions for simulated events are shown for {\sc Diffvm} elastic and proton-dissociative processes 
of $J/\psi(1S)$ (brown hatched histogram) and $\psi(2S)$ (purple hatched histogram) and for a continuous background of muon pairs ({\sc Grape}, green solid histogram) from the 
Bethe--Heitler process. The solid blue histogram represents the sum of all processes.  The relative contribution of different processes was 
obtained from a fit to the data in the range $2 < M(\mu^+\mu^-) < 6$\,GeV.  The result of a double-Gaussian fit to the resonant peaks and 
parameterisation of the background, described in the text, is also shown (solid and dashed lines). 
}
\label{fig:dimuon-mass-w}
\vfill
\end{figure}

\begin{figure}[p]
\vfill
\begin{center}
\includegraphics[trim={0cm 0cm 1cm 0.cm},clip,width=0.49\textwidth]{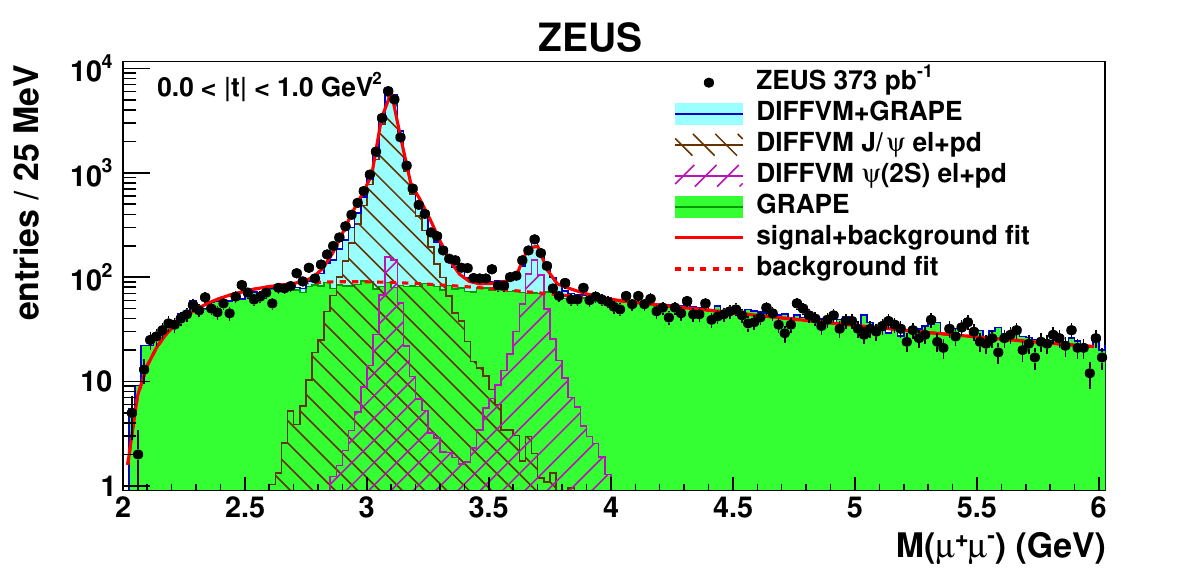}
\put(-185,90){\makebox(0,0)[tl]{(a)}}
\includegraphics[trim={0cm 0cm 1cm 0.cm},clip,width=0.49\textwidth]{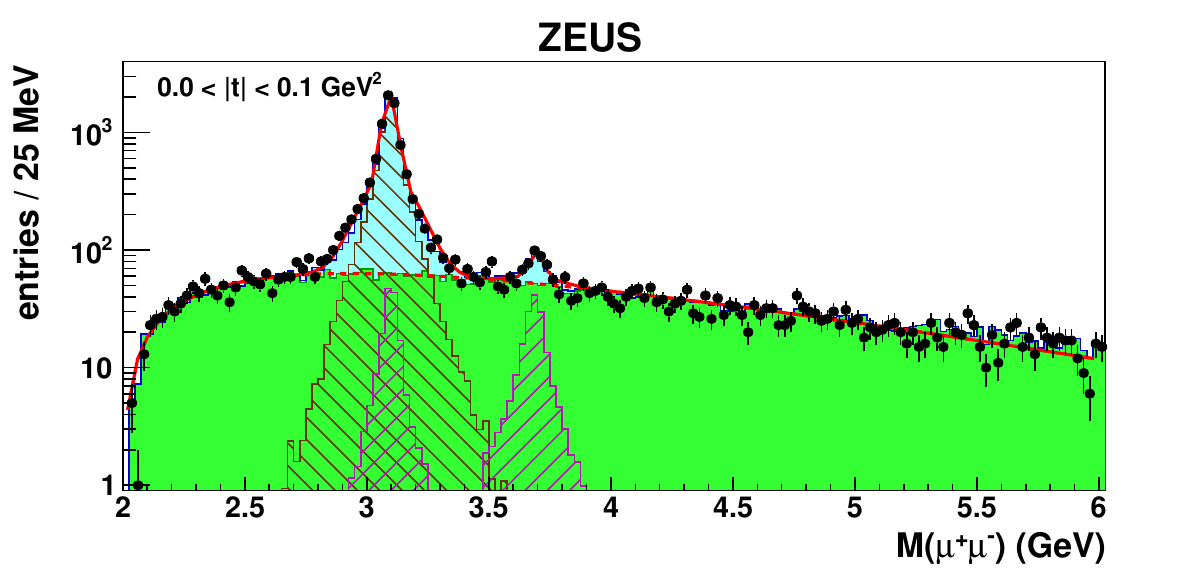}
\put(-185,90){\makebox(0,0)[tl]{(b)}}\\
\includegraphics[trim={0cm 0cm 1cm 0.cm},clip,width=0.49\textwidth]{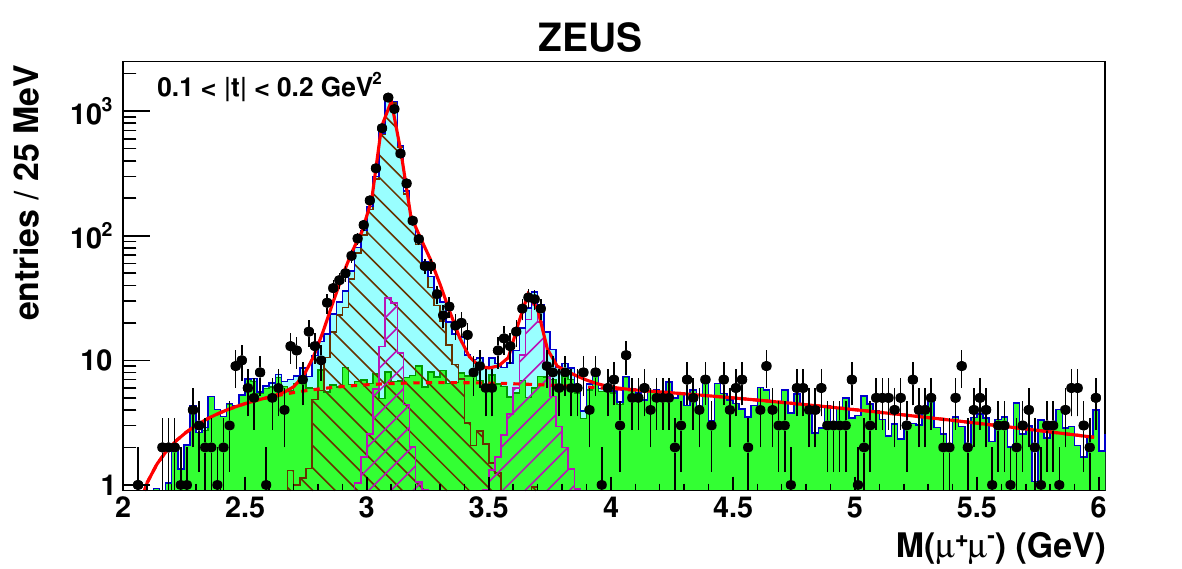}
\put(-185,90){\makebox(0,0)[tl]{(c)}}
\includegraphics[trim={0cm 0cm 1cm 0.cm},clip,width=0.49\textwidth]{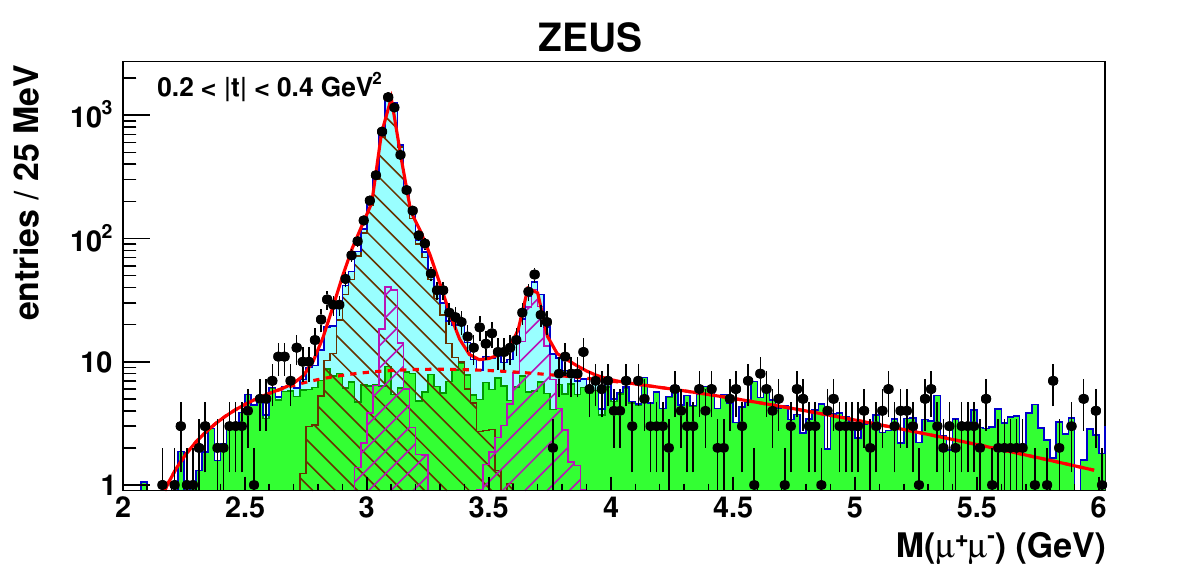}
\put(-185,90){\makebox(0,0)[tl]{(d)}}\\
\includegraphics[trim={0cm 0cm 1cm 0.cm},clip,width=0.49\textwidth]{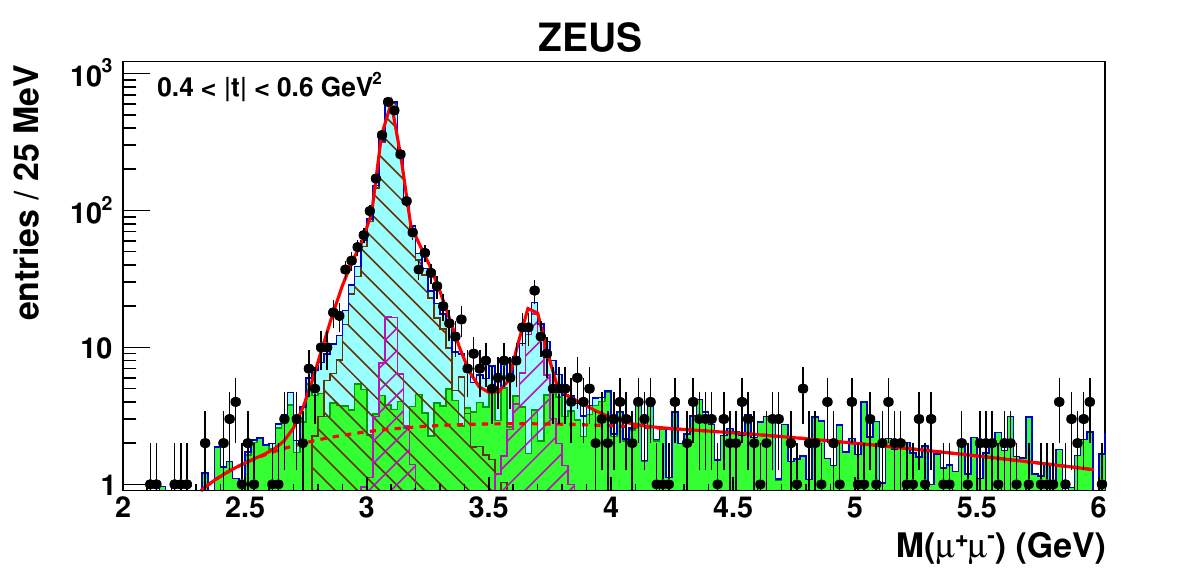}
\put(-185,90){\makebox(0,0)[tl]{(e)}}
\includegraphics[trim={0cm 0cm 1cm 0.cm},clip,width=0.49\textwidth]{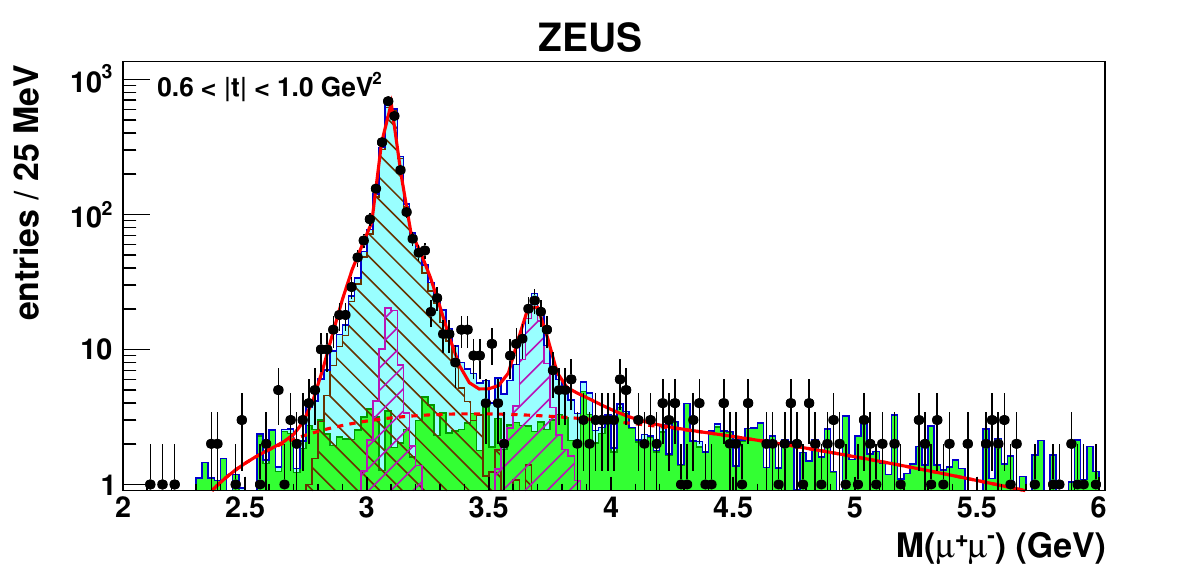}
\put(-185,90){\makebox(0,0)[tl]{(f)}}
\end{center}
\vspace{-0.5cm}
 \caption{
Measured invariant mass distribution, $M(\mu^+\mu^-)$, of dimuon pairs (solid circles) in selected photoproduction events with error bars denoting statistical uncertainties.  
The data are shown in (a) the full $|t|$ range, $0.0 < |t| < 1.0$\,GeV\,$^2$, and (b) -- (f) finer $|t|$ intervals within the full range.  (Note that (a) 
shows the same data and simulations as Fig.~\ref{fig:dimuon-mass-w}(a) but is shown here again to highlight the $|t|$ dependence of the mass 
distributions.)  All further details are as in the caption for Fig.~\ref{fig:dimuon-mass-w}.
}
\label{fig:dimuon-mass-t}
\vfill
\end{figure}

Figure~\ref{fig:dimuon-mass-w} shows the $\mu^+ \mu^-$ mass distributions between 2 and 6\,GeV for the selected events in 
the full region, $30 < W < 180$\,GeV, and in $W$ intervals within this full range in which the 
cross section is measured.   Figure~\ref{fig:dimuon-mass-t} shows the $\mu^+ \mu^-$ mass distribution in the full 
region, $0.0 < |t| < 1.0$\,GeV$^2$, and in $|t|$ intervals within this full range in which the cross section is measured.  Clear 
$J/\psi (1S)$ and $\psi(2S)$ peaks, with masses consistent with those in the PDG~\cite{ptep:083C01}, are seen and no other 
significant peak is observed.

Expectations from MC simulations are also shown in Figs.~\ref{fig:dimuon-mass-w} and~\ref{fig:dimuon-mass-t}, where here and in 
all subsequent figures showing MC simulations, the sum of all the MC distributions is normalised to data.  The relative contribution of 
each different process was obtained from a fit to the data in the range $2 < M(\mu^+\mu^-) < 6$\,GeV.  The 
$J/\psi (1S)$ and $\psi(2S)$ peaks in all 
$W$ and $|t|$ ranges are consistent with events from elastic and proton-dissociative processes.  The varying width of the peak with 
increasing $W$ is due to the different amount of tracking information available; low $W$ and high $W$ corresponds to muons in the 
forward and rear directions respectively where the resolution is less good than in the central tracking region, i.e.\ $60<W<120$\,GeV.  
The different $M(\mu^+ \mu^-)$ resolutions with $W$ are reproduced well by the detector MC simulation.  The distributions outside of the $J/\psi (1S)$ 
and $\psi(2S)$ peaks are, according to MC simulations, consistent with those arising from the Bethe--Heitler process.  The width of the peak 
does not change with varying $|t|$ because $|t|$ is not correlated with the angular distribution of the muons.  However, the Bethe--Heitler 
background decreases significantly with increasing $|t|$.

The numbers of $J/\psi (1S)$ and $\psi(2S)$ mesons were obtained from a fit to the data to describe the peaks and the background.
Each of the peaks was fitted using the sum of two Gaussian functions centred at the same mean value.  The fit 
was further constrained by imposing the same ratios of the widths and the normalisations for the two Gaussian shapes describing the $J/\psi(1S)$ and 
$\psi(2S)$ peaks. This was motivated by the observed scaling of the mass resolution with increasing mass of the resonance and 
stabilises the fit of the $\psi(2S)$ peak, which has a smaller number of candidates.  The background function used was 
$F(x) = A (x - B)^C \cdot \exp(-D[x - B] - E[x - B]^2)$ where $x=M(\mu^+\mu^-)$, $A$, $C$ $D$ and $E$ are parameters determined 
in the fit and $B$ represents the kinematic onset 
of the distribution and is fixed to 2\,GeV, twice the minimum $p_T$ of a muon.  The results of these fits are also shown in Figs.~\ref{fig:dimuon-mass-w} 
and~\ref{fig:dimuon-mass-t} where they describe the data well.  A resonant background in the $J/\psi(1S)$ peak from the decay 
of $\psi(2S)$ mesons where the other decay products are not reconstructed is also shown as part of the $\psi(2S)$ MC distribution; 
this was estimated to be about 2.4\% and will later be subtracted from the Gaussian fit.  The non-resonant background under the $J/\psi(1S)$ peak from the 
Bethe--Heitler process is on average about 9\% of the size of the signal.  Under the $\psi(2S)$ peak, the background from the 
Bethe--Heitler process is about a factor of 2.5~times higher than the signal.  A resonant background under the $\psi(2S)$ peak arises due to leakage from 
the reconstruction of $J/\psi$ mesons, the 
upper tail of which overlaps with the $\psi(2S)$ mass region. It is on average 15\% of the $\psi(2S)$ signal.

\begin{figure}[p]
\vfill
\begin{center}
\includegraphics[trim={0cm 0cm 1.5cm 0.cm},clip,width=0.49\textwidth]{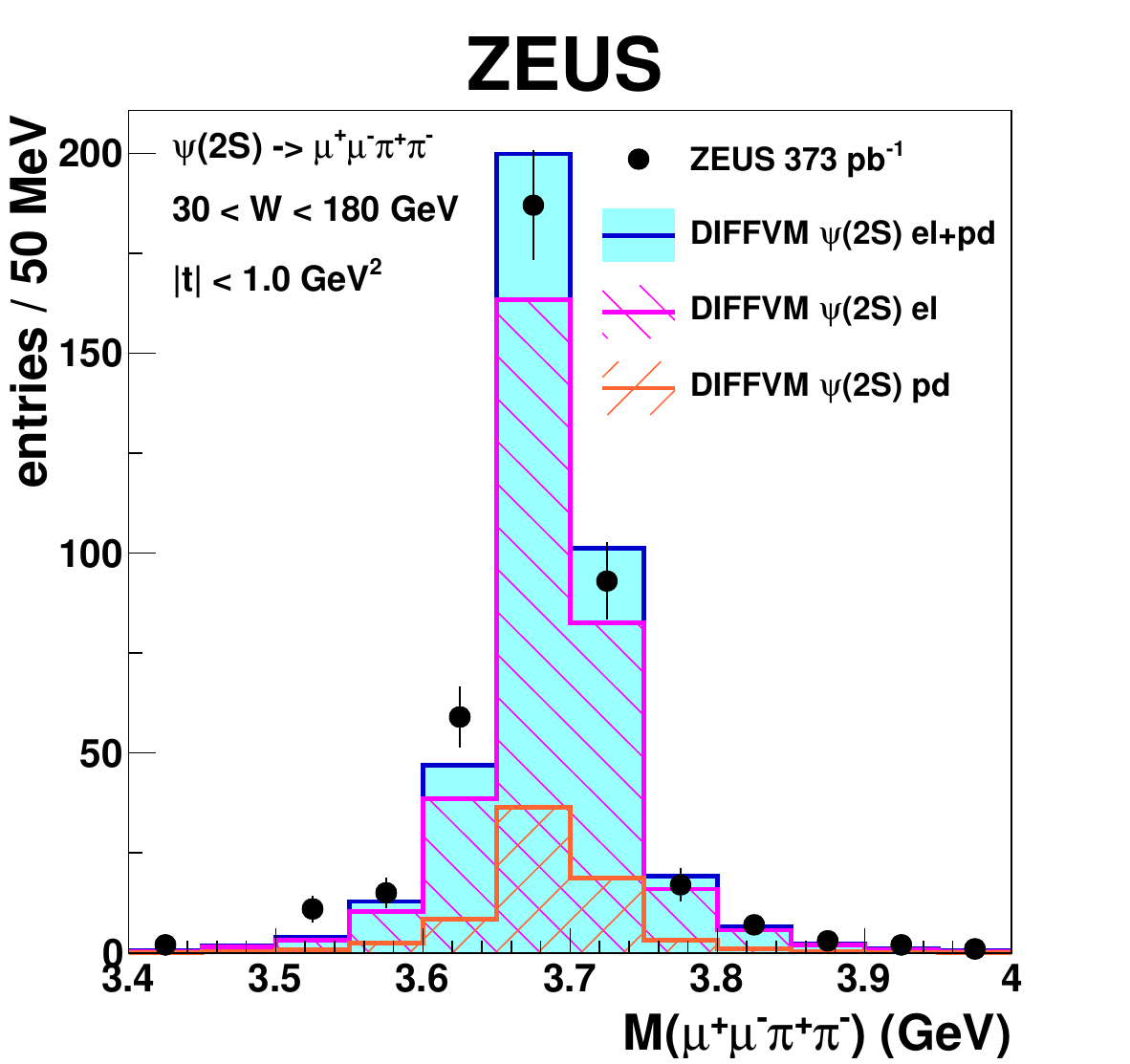}
\put(-185,150){\makebox(0,0)[tl]{(a)}}
\includegraphics[trim={0cm 0cm 1.5cm 0.cm},clip,width=0.49\textwidth]{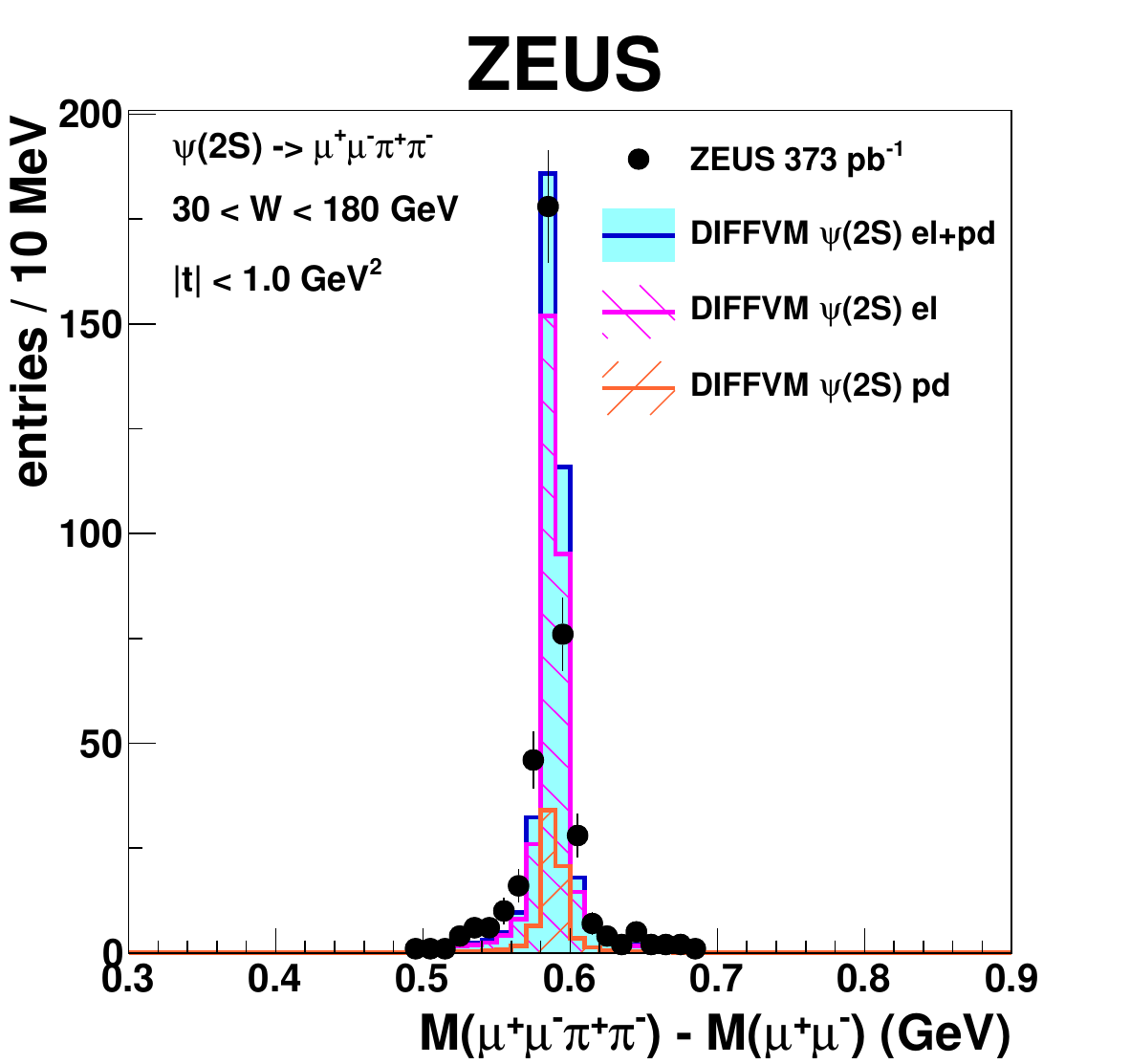}
\put(-185,150){\makebox(0,0)[tl]{(b)}}
\end{center}
 \caption{
(a) Measured invariant mass distribution, $M(\mu^+\mu^-\pi^+\pi^-)$, and (b) difference in invariant masses, 
$M(\mu^+\mu^-\pi^+\pi^-) - M(\mu^+\mu^-)$, for the 4-prong decay of $\psi(2S)$ in photoproduction events (solid circles), with 
error bars denoting statistical uncertainties.  The invariant mass $M(\mu^+\mu^-)$ was required to be in the range 
$2.8 < M(\mu^+\mu^-) < 3.4$ GeV.  Monte Carlo distributions for simulated events generated with {\sc Diffvm} are 
shown for elastic (magenta hatched histogram) and proton-dissociative (orange hatched histogram) processes of 
$\psi(2S)$ as well as their sum (blue solid histogram).   The relative fraction of elastic and proton-dissociative processes 
was determined from a fit to the $|t|$ distribution.
}
\label{fig-psi-mass}
\vfill
\end{figure}

Figure~\ref{fig-psi-mass} shows  the $\mu^+ \mu^- \pi^+ \pi^-$ mass distribution between 3.4 and 4\,GeV and the difference in masses, 
$M(\mu^+ \mu^- \pi^+ \pi^-) - M(\mu^+\mu^-)$, for the selected events, with the additional requirements of $2.8<M(\mu^+\mu^-)<3.4$\,GeV 
and $0.5 < M(\mu^+ \mu^- \pi^+ \pi^-) - M(\mu^+\mu^-) < 0.7$\,GeV.  Clear, narrow peaks are observed in both distributions, especially 
for the mass difference, and both are described well 
by MC simulations.  The distributions are consistent with events from elastic and proton-dissociative processes with a small non-resonant 
background that is about 2 $-$ 3\% of the signal size.  The number of background events in the $\psi(2S) \to \mu^+ \mu^- \pi^+ \pi^-$ sample 
was estimated from data by counting the side-band events in the $M(\mu^+ \mu^- \pi^+ \pi^-) - M(\mu^+\mu^-)$ distribution outside the signal 
region, before applying the requirement on this quantity. The background events were counted in the $0.7-1.5$\,GeV interval and the obtained number 
was rescaled to the signal interval $0.5 - 0.7$\,GeV (see Fig.~\ref{fig-psi-mass}(b)) assuming a uniform distribution.  Such a procedure was 
performed for each $W$ and $|t|$ interval.  The number of $\psi(2S)$ mesons was found by counting the number of entries and subtracting the 
background in the range $3.4<M(\mu^+ \mu^- \pi^+ \pi^-)<4$\,GeV.

The numbers of signal and background events and their statistical uncertainties used for further analysis are given in Table~\ref{tab:tab1} 
for each channel and in five $W$ and five $|t|$ regions.  

\subsection{Correction procedure and comparison of measured and simulated distributions}
\label{sec:data-MC}

In order to determine the acceptance using simulated events,  simulated and measured distributions have to agree.  To achieve this, 
corrections for the efficiency of muon reconstruction were developed for the MC simulation.  Muon-identification corrections 
were developed using a sample of exclusive dimuon events in which one muon was tagged and the probability of 
reconstruction of the other muon 
evaluated.  The probability was determined for the full reconstruction chain, including the trigger efficiency, in bins of $\eta^\mu$ and 
$p_Z^\mu/p_T^\mu/p^\mu$, depending on whether the muon was reconstructed in the R/B/FMUON.  This leads to typical 
efficiencies of $20-40\%$, averaged over $\eta^\mu$ and momentum, although in individual bins of $\eta^\mu$ and 
$p_Z^\mu/p_T^\mu/p^\mu$ these can be under 10\%, mainly at low momentum where the muons do not reach the detectors.  They can 
reach up to 60\% at high $p_Z^\mu$ and high $p_T^\mu$.  The efficiency to 
reconstruct muons in the CAL was typically above 90\%.  After application of the data-driven corrections, the data and MC distributions 
agree well.

The CTD first-level trigger used in the selection of events has an efficiency that depends on the track multiplicity and needs to be 
evaluated for the $\mu^+ \mu^- \pi^+ \pi^-$ final state with an independent trigger.  A sample of DIS events~\cite{np:b909:934}, 
passing an independent trigger chain but with the same final state, was used to determine this correction.  To ensure the same tracking 
topology, the scattered electron was restricted to the RCAL cells close the beam-pipe with no matched track.  This correction 
was consistent with unity to within about $\pm 5\%$.

The tracking efficiency for low-momentum pions ($p_T < 0.26$\,GeV) is overestimated in MC simulations and so a correction was 
applied~\cite{thesis:Bachynska:2012,Libov:2012nlc} in simulations to $\psi(2S)$ decays to $\mu^+\mu^-\pi^+\pi^-$.  An 
event was assigned a weight given by $w = 1 + 0.548 \cdot (p_T^\pi - 0.26)$ if one pion had transverse momentum, $p_T^\pi$, below 
0.26\,GeV.  If both pions had transverse momentum below 0.26\,GeV, the quantity $w$ was calculated for each pion and the event 
weighted by the product of the two weights.

\begin{figure}[p]
\vfill
\begin{center}
\includegraphics[trim={0cm 0cm 1.5cm 0.cm},clip,width=0.49\textwidth]{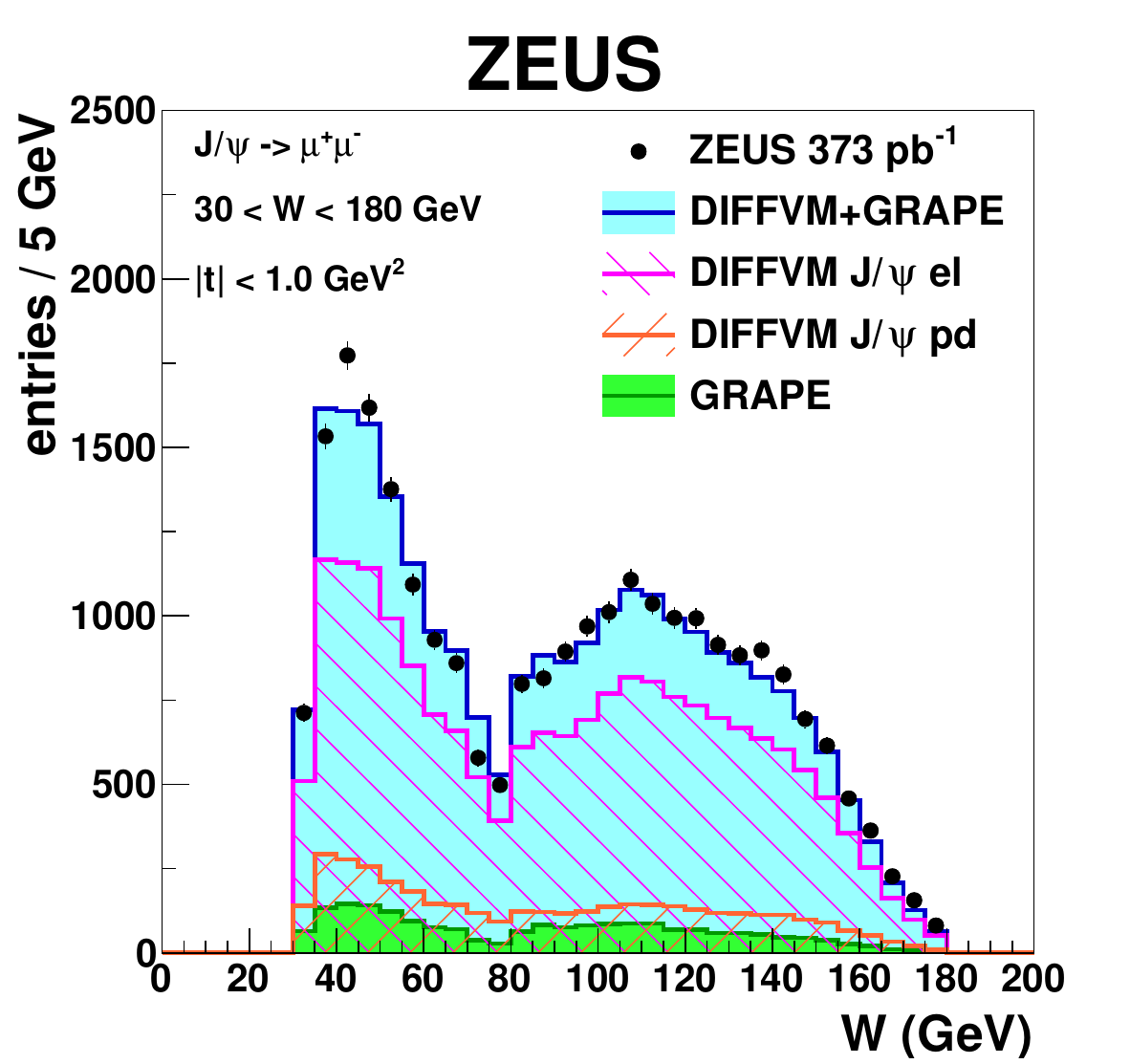}
\put(-35,120){\makebox(0,0)[tl]{(a)}}
\includegraphics[trim={0cm 0cm 1.5cm 0.cm},clip,width=0.49\textwidth]{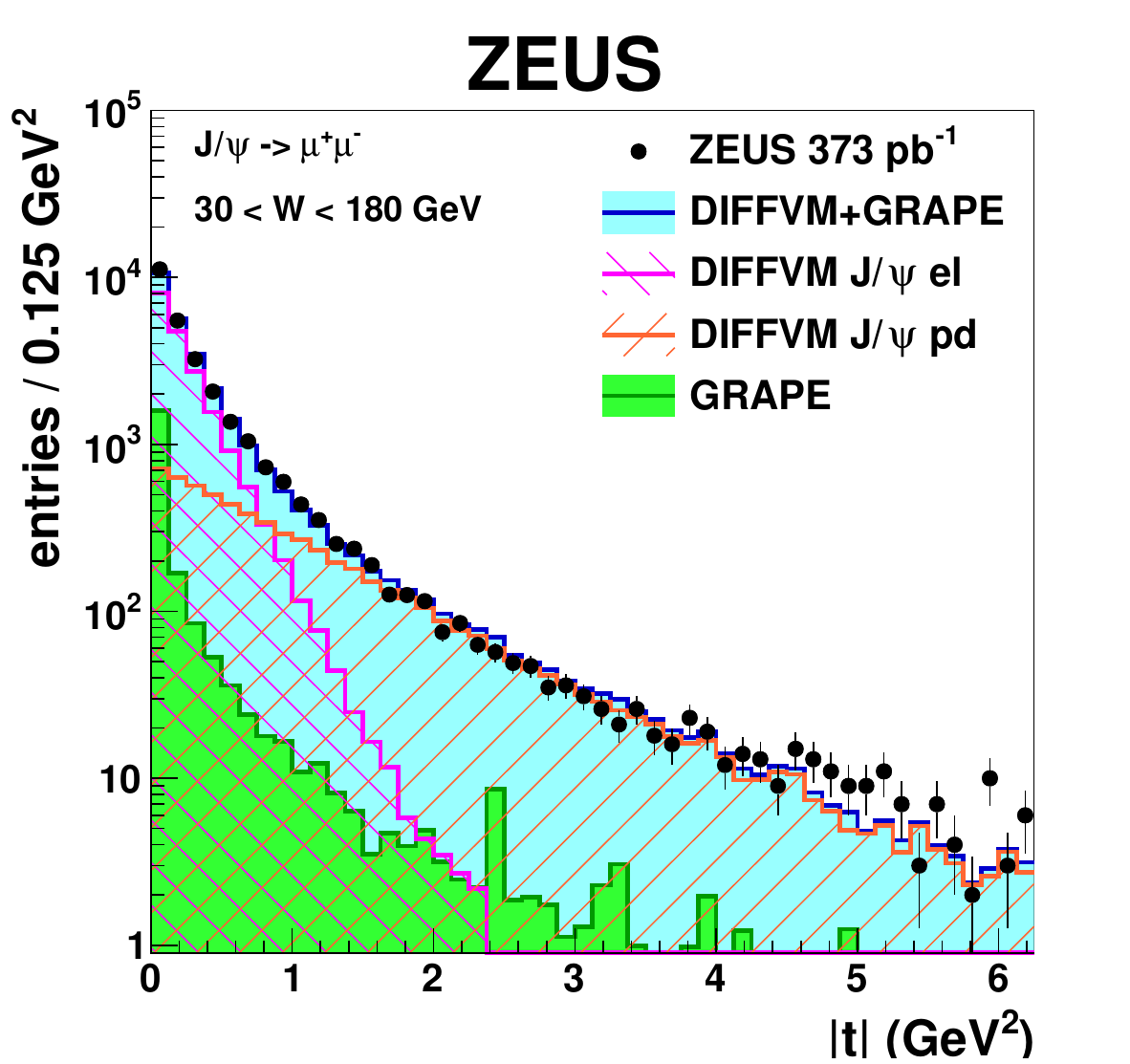}
\put(-35,120){\makebox(0,0)[tl]{(b)}}\\
\includegraphics[trim={0cm 0cm 1.5cm 0.cm},clip,width=0.49\textwidth]{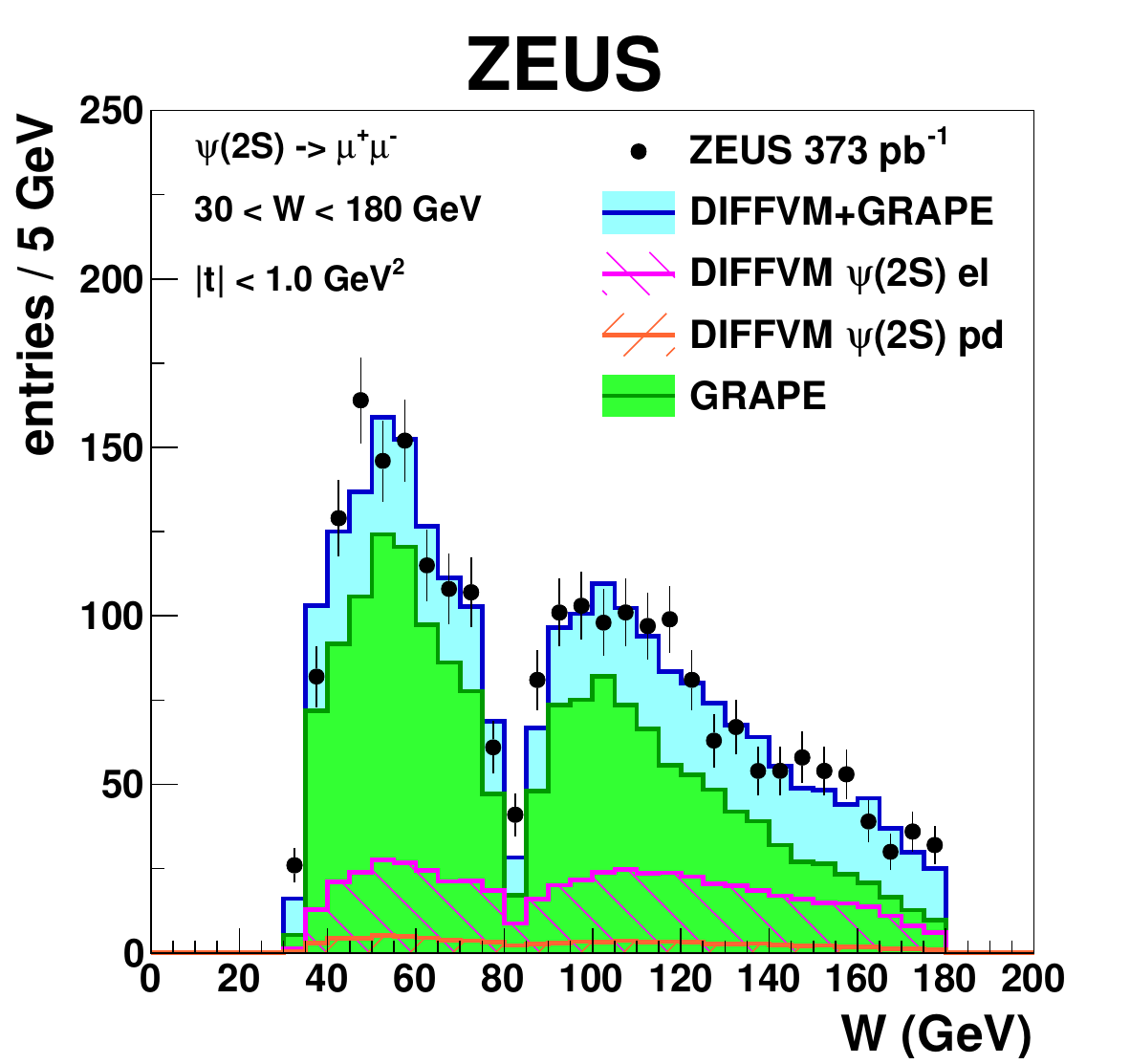}
\put(-35,120){\makebox(0,0)[tl]{(c)}}
\includegraphics[trim={0cm 0cm 1.5cm 0.cm},clip,width=0.49\textwidth]{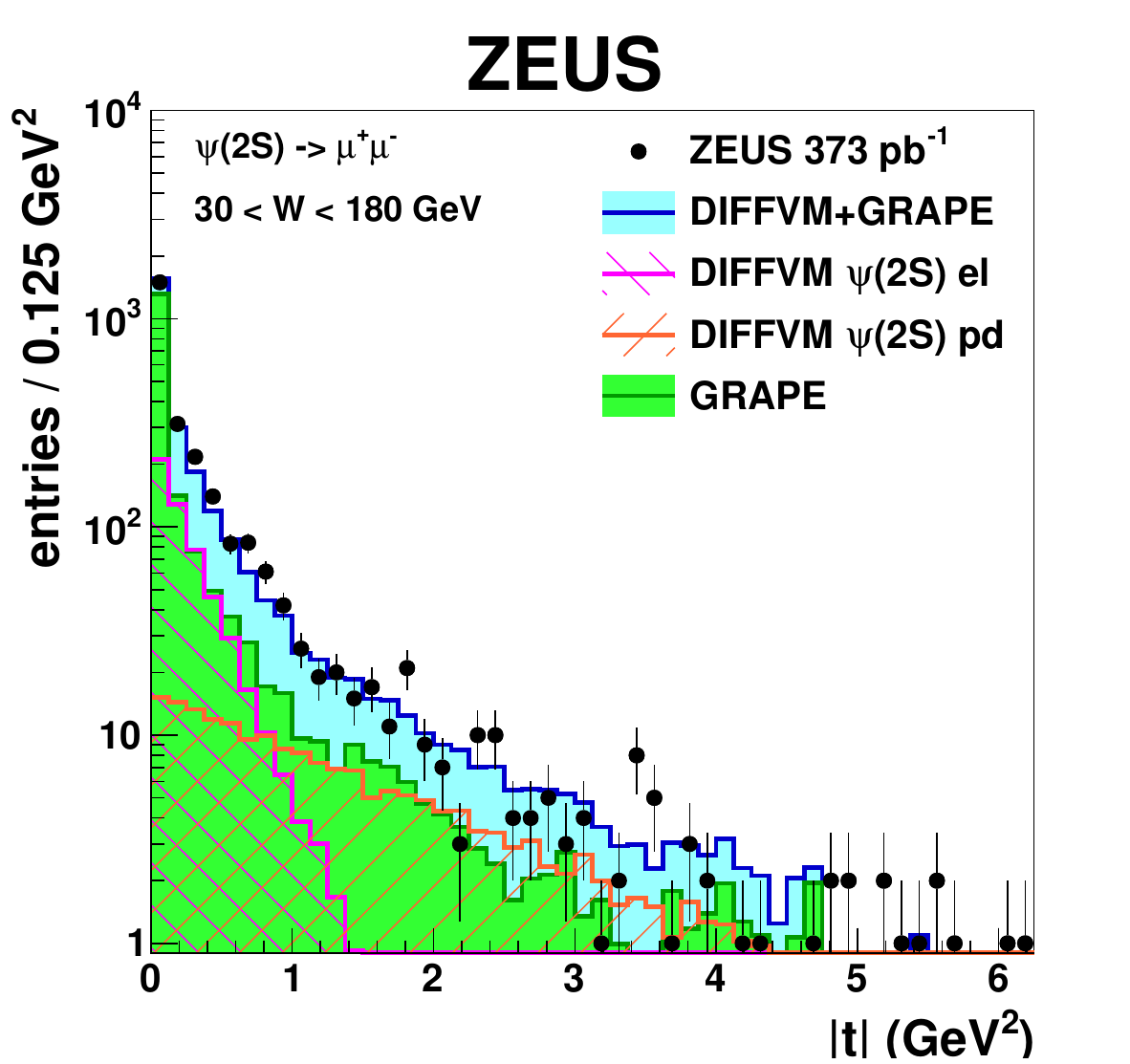}
\put(-35,120){\makebox(0,0)[tl]{(d)}}\\
\end{center}
 \caption{
Distributions of (a, c) $W$ and (b, d) $|t|$ reconstructed for the decay of (a, b) $J/\psi(1S)$ and (c, d) $\psi(2S)$ mesons to a $\mu^+\mu^-$ pair 
in photoproduction events (solid circles), with error bars denoting statistical uncertainties.   {\sc Diffvm} MC distributions for simulated events are shown for 
elastic (magenta hatched histogram) and proton-dissociative (orange hatched histogram) processes separately.  The total sum of all contributions 
(blue solid histogram) as well as the {\sc Grape} MC contribution from Bethe--Heitler processes (green solid histogram) is also shown.   The statistical uncertainties on the MC histograms are not shown; they are about 40\% of the size of the statistical uncertainties in the data.
}
\label{fig-controlplots1}
\vfill
\end{figure}

Data and MC simulations with a $\mu^+\mu^-$ pair are compared in Fig.~\ref{fig-controlplots1} for $2.8 < M(\mu^+ \mu^-) < 3.4$\,GeV 
and $3.4 < M(\mu^+ \mu^-) < 4.0$\,GeV, corresponding to the mass ranges of the $J/\psi(1S)$ and $\psi(2S)$, respectively, after 
application of all corrections discussed above.  The structures in the $W$ data 
distribution reflect the acceptance of the detector, in particular that of the muon detectors, with the dips around 80\,GeV due to the requirements to remove 
cosmic-ray events.  The data distributions are well described by MC simulations.  
This demonstrates the validity of the correction procedure for the muon acceptance and tuning of the MC simulation parameters.  The data distribution in $|t|$ exhibits an 
exponential fall-off with increasing $|t|$ and is well described by the mixture of MC samples.  The fraction of proton-dissociative 
events increases significantly with increasing $|t|$, becoming the dominant process above 1\,GeV$^2$; this justifies the requirement 
in the analysis of $|t| < 1$\,GeV$^2$ in order to enrich the sample in elastic events.

\begin{figure}[p]
\vfill
\begin{center}
\includegraphics[trim={0cm 0cm 1.5cm 0.cm},clip,width=0.49\textwidth]{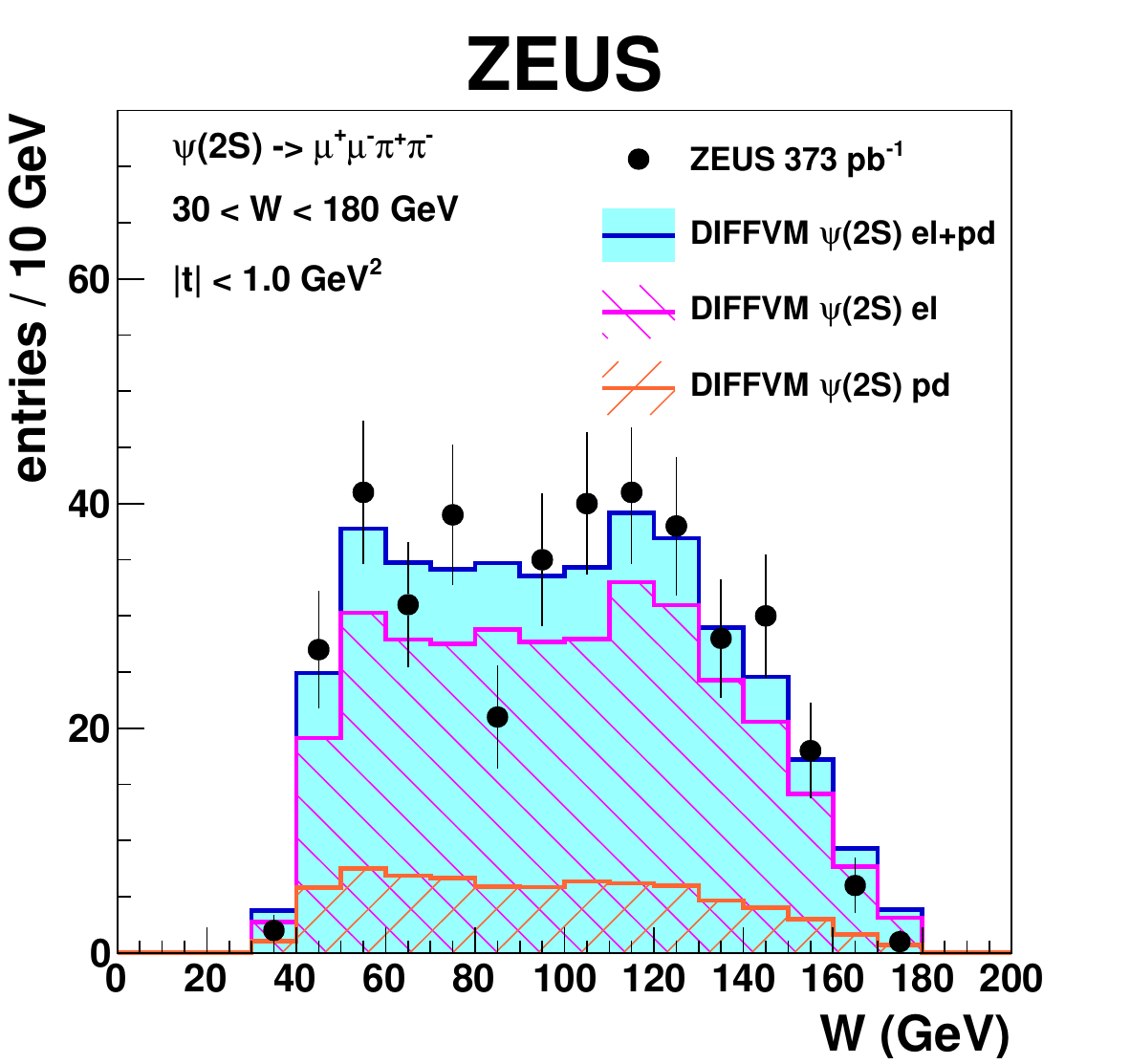}
\put(-40,130){\makebox(0,0)[tl]{(a)}}
\includegraphics[trim={0cm 0cm 1.5cm 0.cm},clip,width=0.49\textwidth]{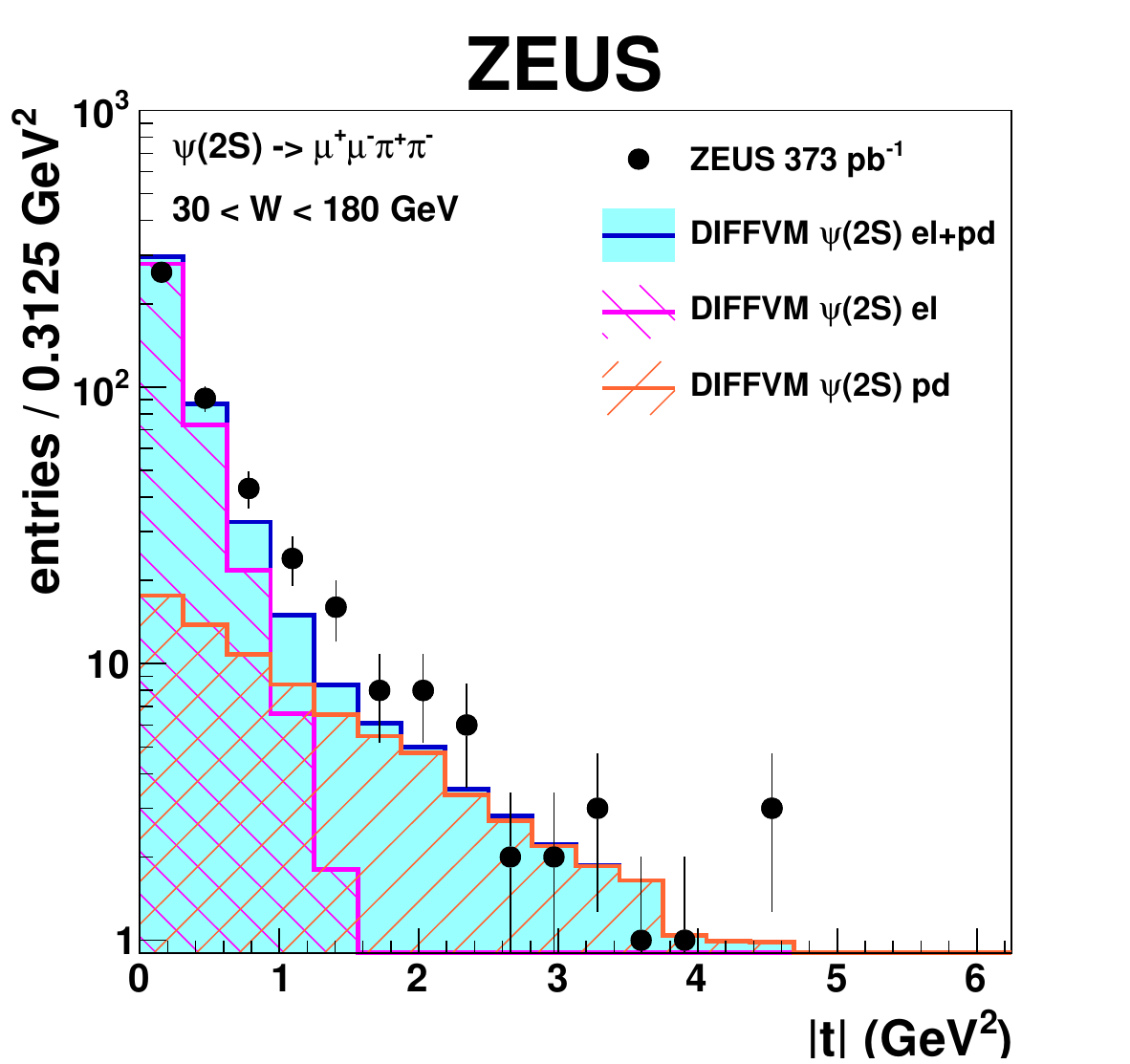}
\put(-40,130){\makebox(0,0)[tl]{(b)}}
\end{center}
 \caption{
Distributions of (a) $W$ and (b) $|t|$ reconstructed for the 4-prong decay of $\psi(2S)$ in photoproduction events (solid circles), 
with error bars denoting statistical uncertainties.  
Details of the MC distributions for the 4-prong decay are as in the caption for Fig.~\ref{fig-psi-mass}.
}
\label{fig-controlplots2}
\vfill
\end{figure}

Figure~\ref{fig-controlplots2} shows distributions in $W$ and $|t|$ when $\mu^+\mu^-\pi^+\pi^-$ were observed in the 
final state.  The $W$ distribution is reasonably flat in the range $40 < W < 150$\,GeV with fall-offs either side of this region.  
The $|t|$ distribution exhibits an exponential fall-off with increasing $|t|$.  The MC simulations give a reasonable description of the 
data.


\section{Cross-section ratio $\boldsymbol{\psi (2S)}$ to $\boldsymbol{J/\psi (1S)}$}
\label{ratio}

The following cross-section ratios, $\sigma_{\psi(2S)} / \sigma_{J/\psi (1S)}$, have been measured:  
$R_{\mu \mu}$ for $\psi (2S ) \rightarrow \mu ^+ \mu^-$,  $R_{J/\psi \,\pi \pi} $ for $\psi (2S ) \rightarrow J/\psi (1S ) \,\pi^+ \pi^-$ 
and $R$ for the combination of the two decay modes.  In each case, the decay $J/\psi (1S) \to \mu ^+ \mu^-$ was 
used in the denominator.

\subsection{Determination of the cross-section ratio}
\label{sect:ratio}

The cross-section ratios for each bin and the full sample were calculated using
$$ R_{\mu \mu} =
   \left[ \left( \frac{N_{\mu \mu}^{\psi (2S)}} {B_{\psi (2S) \to \mu ^+ \mu^-} \cdot A_{\mu \mu}^{\psi (2S)} } \right)\Big/
   \left( \frac{N_{\mu \mu}^{J/\psi(1S)}} {B_{J/\psi (1S) \to \mu ^+ \mu^-} \cdot A_{\mu \mu}^{J/\psi(1S)}} \right) \right]
   \cdot \frac{1- f_{\rm pdiss}^{\psi (2S)}}{1- f_{\rm pdiss}^{J/\psi (1S)}}
$$

and

$$ R_{J/\psi \, \pi \pi} =
   \left[ \left(\frac{N_{J/\psi \,\pi \pi}^{\psi (2S)}} {B_{\psi (2S ) \to J/\psi (1S ) \,\pi^+ \pi^-} \cdot A_{J/\psi \, \pi \pi}^{\psi (2S)}} \right) \Big/
   \left(\frac{N_{\mu \mu}^{J/\psi(1S)}} {A_{\mu \mu}^{J/\psi(1S)}} \right) \right] 
   \cdot \frac{1- f_{\rm pdiss}^{\psi (2S)}}{1- f_{\rm pdiss}^{J/\psi (1S)}}
   \,. $$

Here $N_i^j$ denotes the number of observed signal events for the charmonium state $j$ with the decay mode $i$, $A_i^j$ is 
the corresponding acceptance determined from the ratio of reconstructed to generated MC events after reweighting, and 
$f_{\rm pdiss}^j$ is the fraction of proton-dissociative events.  
The  value of $f_{\rm pdiss}^j$ was determined by fitting 
the $|t|$ distribution, for $|t| < 6.25\,{\rm GeV}^2$, of the data with the $|t|$ distributions from MC samples.  
It was found that $f_{\rm pdiss}^j$ is independent of $W$ and so the 
mean values of  $f_{\rm pdiss}^{J/\psi (1S)} = 17\%$ and $f_{\rm pdiss}^{\psi (2S)} = 16\%$ were used for the determination of $R$ 
as a function of $W$.  The value of $f_{\rm pdiss}^j$ has a 
strong dependence on $|t|$, varying from about 7\% for $0 < |t| < 0.1$\,GeV$^2$ to 45\% for $0.6 < |t| < 1$\,GeV$^2$ (see 
Table~\ref{tab:tab1} for more details). The corresponding values in each bin were used in the determination of $R$ as a function of $|t|$.
However, there is little difference between $f_{\rm pdiss}^{J/\psi (1S)} $ and $f_{\rm pdiss}^{\psi (2S)}$  and so the final 
factor in the calculation of $R_{\mu \mu}$ and $R_{J/\psi \, \pi \pi}$ is approximately unity.  The following 
values were used for the branching fractions:  $B_{J/\psi (1S) \to \mu ^+ \mu^-} = (5.961 \pm 0.033 )\%$,  
$B_{\psi (2S) \to \mu ^+ \mu^-} = (0.80 \pm 0.06 )\%$ and 
$B_{\psi (2S ) \to   \mu ^+ \mu^- \pi^+ \pi^-} = (2.07 \pm 0.02 )\%$~\cite{ptep:083C01}.

\begin{figure}[p]
\vfill
\begin{center}
\includegraphics[trim={0cm 0cm 1.25cm 0.cm},clip,width=0.6\textwidth]{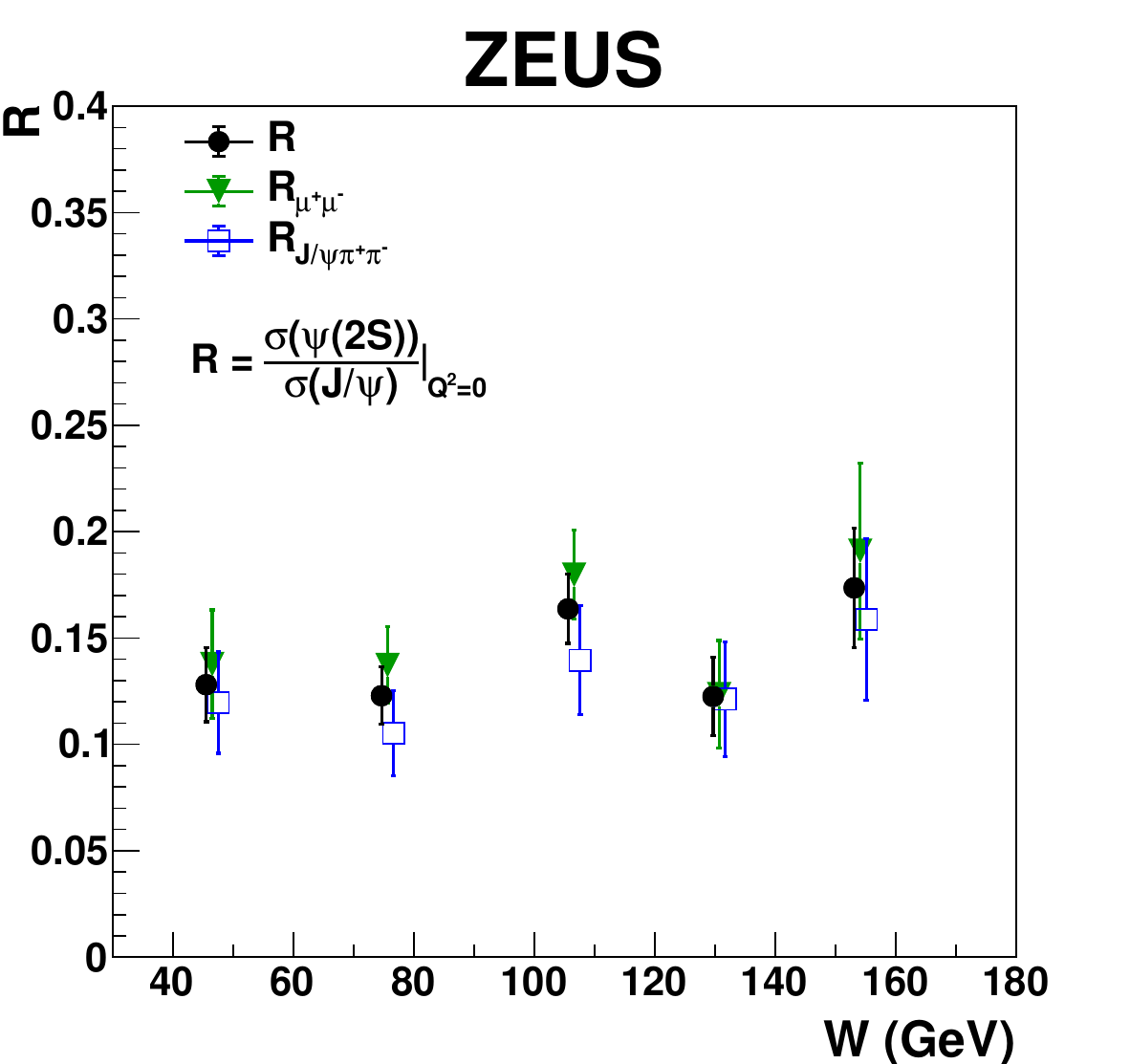}
\put(-40,230){\makebox(0,0)[tl]{(a)}}\\
\includegraphics[trim={0cm 0cm 1.25cm 0.cm},clip,width=0.6\textwidth]{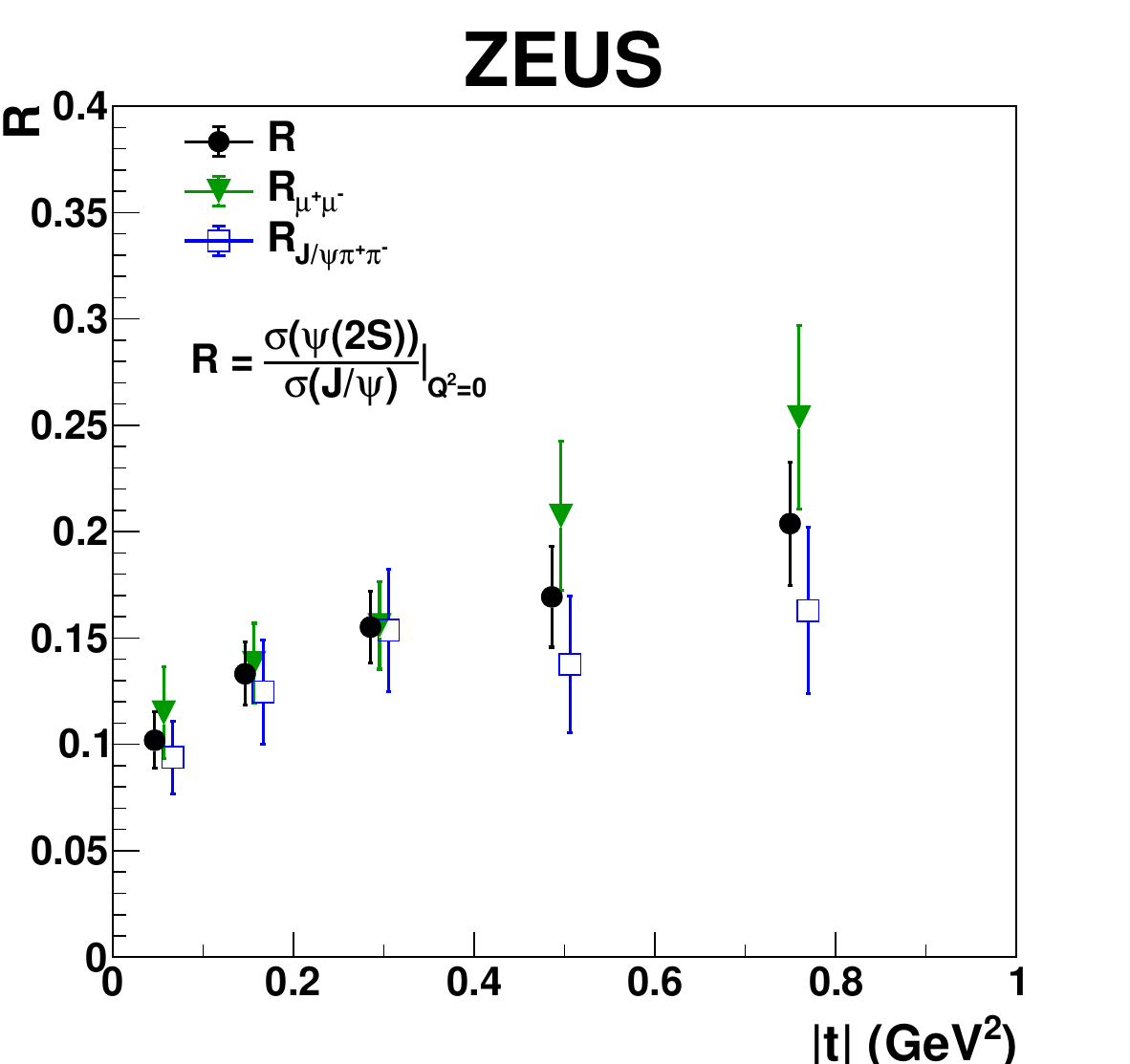}
\put(-40,230){\makebox(0,0)[tl]{(b)}}\\
\end{center}
 \caption{
Cross-section ratio $R = \sigma_{\psi(2S)}/\sigma_{J/\psi(1S)}$ in photoproduction as a function of (a) $W$ and (b) $|t|$ for the two 
decay channels, $R_{\mu \mu}$ for 
$\psi (2S ) \rightarrow \mu ^+ \mu^-$ (triangles) and $R_{J/\psi \,\pi \pi} $ for $\psi (2S ) \rightarrow J/\psi (1S ) \,\pi^+ \pi^-$ 
(squares), and the combination of the two decay modes (solid circles).  The error bars show the statistical uncertainties 
only.  The points for $R$ are shown at the mean $W$ and $|t|$ values for each bin as determined for the $J/\psi(1S)$ data 
(see Table~\ref{tab:tab1}).  The points for $R_{\mu \mu}$ and $R_{J/\psi \,\pi \pi}$ are displaced horizontally for better visibility. 
}
\label{fig:R-each-comb}
\vfill
\end{figure}

The cross-section ratios for the two decay channels, $R_{\mu \mu}$ for $\psi (2S ) \rightarrow \mu ^+ \mu^-$ and $R_{J/\psi \,\pi \pi} $ 
for $\psi (2S ) \rightarrow J/\psi (1S ) \,\pi^+ \pi^-$ are shown in Fig.~\ref{fig:R-each-comb} in bins of $W$ and $|t|$, with statistical uncertainties only.  
The values are consistent for the two channels.  The two independent measurements of the cross-section ratio  
$\sigma_{\psi(2S)} / \sigma_{J/\psi (1S)}$ were combined.  The combined cross-section ratio, $R$, 
was obtained using the weighted average of the cross sections determined for the two $\psi(2S)$ decay modes. The 
statistical uncertainties were used for the weights. 
The combined cross-section ratio, $R$, is also shown in Fig.~\ref{fig:R-each-comb}, with statistical uncertainties only.

\subsection{Systematic uncertainties}
    \label{sect:Systematics}

The systematic uncertainties on the $R$ values were obtained by performing a suitable variation to determine the change of $R$ 
relative to its central value for each source of uncertainty.  The following sources of systematic uncertainty were 
considered, with typical values given for the change on the final measured $R$ value (see Table~\ref{tab:syst} for full details):

\begin{itemize}

\item the $t$ dependence (exp($-b|t|$)) of the {\sc Diffvm} MC simulations was varied by the uncertainty on the $b$ values: 
$4.6 \pm 0.3$\,GeV$^{-2}$ ($\Delta_1$) and $4.3 \pm 0.7$\,GeV$^{-2}$ ($\Delta_2$) for $J/\psi(1S)$ and $\psi(2S)$ elastic 
events and $1.0 \pm 0.1$\,GeV$^{-2}$ ($\Delta_3$) and $0.7 \pm 0.2$\,GeV$^{-2}$ ($\Delta_4$) for $J/\psi(1S)$ and $\psi(2S)$ proton-dissociative 
events.  Additionally, the $\alpha^\prime$ parameter in the re-weighting of $b$ was varied by its uncertainty, 
$\alpha^\prime = 0.12 \pm 0.04$\,GeV$^{-2}$ ($\Delta_5$).  
The variation in $b$ for $\psi(2S)$ proton-dissociative events led to changes in $R$ that increased with increasing $|t|$, with 
an average change of $\pm 0.01$ in $R$.
The other variations led to typical changes of below $\pm 0.005$ in $R$;

\item the $W$ dependence ($W^\delta$) of the {\sc Diffvm} MC simulations was varied by the uncertainty on $\delta$ values: 
$0.67 \pm 0.10$ ($\Delta_6$) and $1.10 \pm 0.20$ ($\Delta_7$) for $J/\psi(1S)$ and $\psi(2S)$ elastic events and 
$0.42 \pm 0.15$ ($\Delta_8$) and $0.70 \pm 0.30$ ($\Delta_9$) for $J/\psi(1S)$ and $\psi(2S)$ proton-dissociative events.  Typical variations were 
$\pm 0.001$ in $R$;

\item the $M_Y$ dependence ($1/M_Y^\beta$) of the {\sc Diffvm} MC simulations was varied by an uncertainty on $\beta$, 
$2.4 \pm 0.3$ ($\Delta_{10}$ and $\Delta_{11}$), estimated from comparisons to previous H1 and ZEUS analyses.  These variations 
led to a change of less than $\pm 0.001$ in $R$;

\item the correction factors determined in $\eta^\mu$ and $p_Z^\mu/p_T^\mu/p^\mu$ bins for the muon efficiencies were varied 
by doubling the bin size in the $\eta^\mu$ and $p_Z^\mu/p_T^\mu/p^\mu$ grid ($\Delta_{12}$).  These variations led to a change of 
$-0.001$ in $R$;

\item the minimum muon-$p_T$ requirement was varied from 1.0\,GeV by $\pm 0.1$\,GeV ($\Delta_{13}$) to check the stability of the 
background estimation 
from the fit at the lower edge of the dimuon mass spectrum and led to a change of $^{+0.002}_{-0.000}$ in $R$;

\item the minimum pion-$p_T$ requirement was varied from 0.12\,GeV by $\pm 0.02$\,GeV ($\Delta_{14}$) , where the value 0.1\,GeV 
is consistent with the 
lower edge of the tracker acceptance, and led to a change of $^{-0.001}_{+0.003}$ in $R$;

\item the pion-candidate tracks in the $\psi(2S) \to \mu^+ \mu^- \pi^+ \pi^-$ decay were required to be associated to the vertex 
rather than the default of no vertex requirement ($\Delta_{15}$).  This led to a change of $-0.007$ in $R$;

\item the transverse-momentum requirement for the correction of pion-candidate tracks was varied from 0.26\,GeV by $\pm 0.04$\,GeV 
($\Delta_{16}$) and led to a 
change of $^{-0.001}_{+0.002}$ in $R$;

\item the maximum energy of a CAL cluster not associated with a muon or pion candidate was varied from the default 0.5\,GeV by 
$\pm 0.1$\,GeV~\cite{np:b909:934} ($\Delta_{17}$) and led to a change of $^{+0.002}_{-0.004}$ in $R$;

\item the maximum allowed energy inside a cone of maximum angle surrounding the FCAL beam-pipe hole used to suppress proton-dissociative events 
were varied from the defaults: $\theta = 0.12 \pm 0.02$\,rad ($\Delta_{18}$) and energy of $1.00 \pm 0.25$\,GeV ($\Delta_{19}$).  All 
variations led to a change of less than or equal to $\pm 0.001$ in $R$;

\item the requirement on the timing difference in the CAL, $t_{\rm down}-t_{\rm up}$ was varied from the default 8\,ns by $\pm 1$\,ns 
($\Delta_{20}$), according to a study of cosmic-ray muons, and led to a change of less than $^{-0.000}_{+0.002}$  in $R$;

\item the numbers of $J/\psi(1S)$ and $\psi(2S)$ mesons were extracted using a MC template fit ($\Delta_{21}$), rather than the two 
Gaussian and background functions, as a check of modelling the background (see Figs.~\ref{fig:dimuon-mass-w} 
and~\ref{fig:dimuon-mass-t}), and led to a change of $-0.010$ in $R$;

\item the branching ratios were varied according to their uncertainties given in Section~\ref{sect:ratio} and led to changes of 
$\pm 0.001$, $\mp 0.007$ and less than $\mp 0.001$ in $R$ for the variations in $B_{J/\psi (1S) \to \mu ^+ \mu^-}$ ($\Delta_{22}$), 
$B_{\psi (2S) \to \mu ^+ \mu^-}$ ($\Delta_{23}$) and $B_{\psi (2S ) \to   \mu ^+ \mu^- \pi^+ \pi^-}$ ($\Delta_{24}$), respectively.

 \end{itemize}

The largest uncertainties arose from the change in the $b$ slope, especially for $\psi(2S)$ proton-dissociative events and especially 
at high $|t|$, the association to the vertex of the pion-candidate tracks, the method for extracting the number of signal events, and the branching ratio 
$B_{\psi (2S) \to \mu ^+ \mu^-}$.  The total systematic uncertainty, given in Table~\ref{tab:tab1}, was obtained from the separate quadratic sums of the positive 
and negative changes in each bin.

A steepening of the $|t|$ distribution to low $M_Y$ has been observed in hadron--hadron diffraction~\cite{pr:d14:3148,np:b108:1}.   To investigate 
this possibility in photoproduction, the $b$ values in the MC simulation for the proton-dissociative events were changed to those extracted from elastic events, 
i.e.\ from 1.0 to 4.6\,GeV$^{-2}$ for $J/\psi(1S)$ and from 0.7 to 4.3\,GeV$^{-2}$ for $\psi(2S)$ events for $M_Y < 1.9$\,GeV.  This 
led to an average change of $-0.009$ in $R$ with a change of $-0.005$ at lowest $|t|$ and $-0.015$ at highest $|t|$.  This was not 
included in the total systematic uncertainty as such a change led to a poor description of the forward energy flow, estimated by the 
sum of energy in the FCAL surrounding the beam-pipe hole ($\theta<0.12$\,rad).


\section{Results}
      \label{sect:Results}

The cross-section ratio $R=\sigma _{\psi(2S )} / \sigma _{J/\psi (1S )}$ has been measured in exclusive photoproduction in the 
kinematic range $Q^2 < 1$\,GeV$^2$, $30 < W < 180$\,GeV and $|t| < 1$\,GeV$^2$ using a total integrated luminosity of 
373~pb$^{-1}$.  The measured value is

\[
R = 0.146 \pm 0.010 \,{\rm (stat.)}\,^{+0.016}_{-0.020} \,{\rm (syst.)}\,,
\]

where the first uncertainty is statistical and the second is the sum of all systematic uncertainties added in quadrature.  This value, well below 1, confirms 
the expected suppression of the $\psi(2S)$ cross section relative to the $J/\psi(1S)$ cross section.

\begin{figure}[p]
\vfill
\begin{center}
\includegraphics[width=0.6\textwidth]{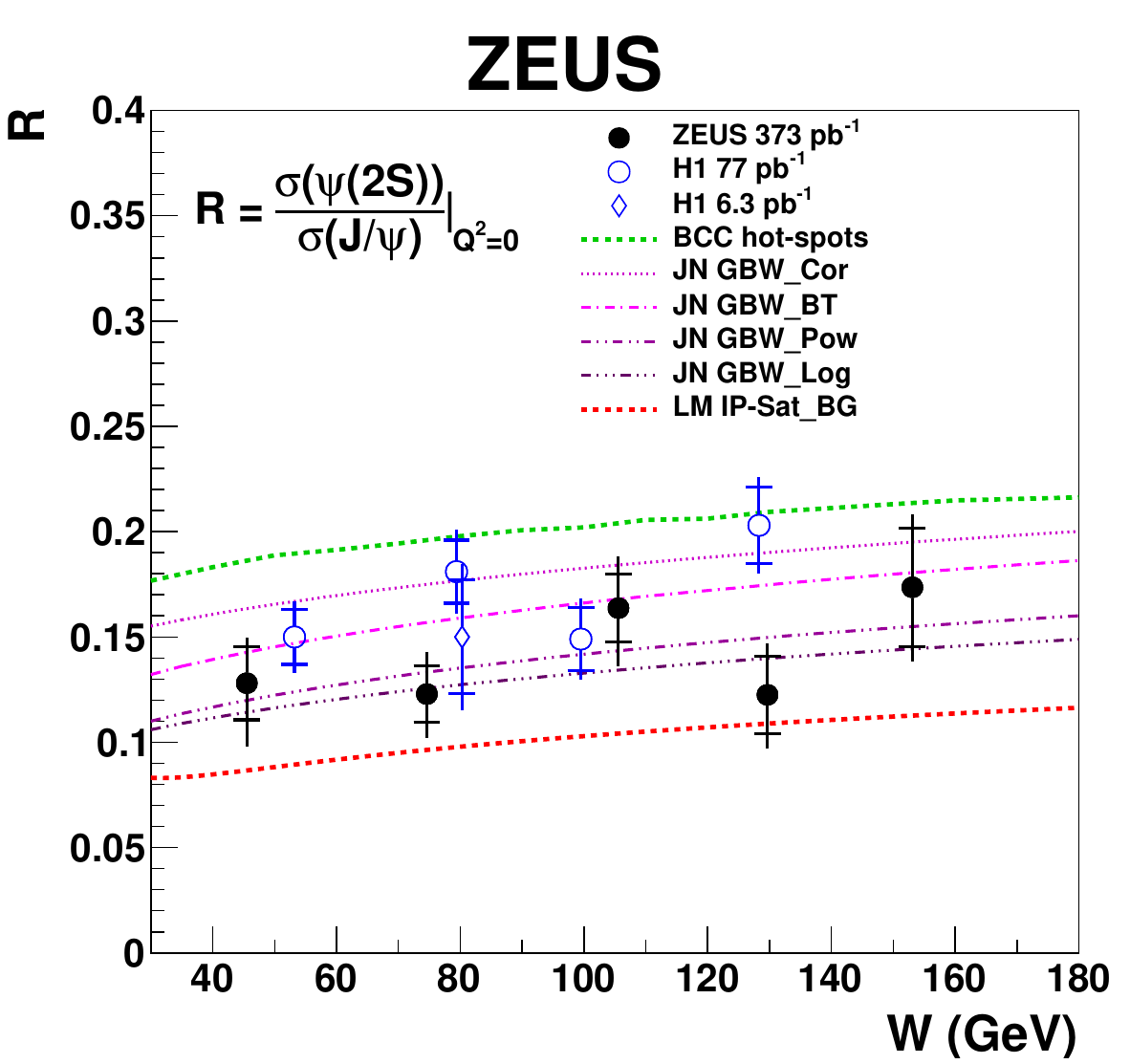}
\put(-40,220){\makebox(0,0)[tl]{(a)}}\\
\includegraphics[width=0.6\textwidth]{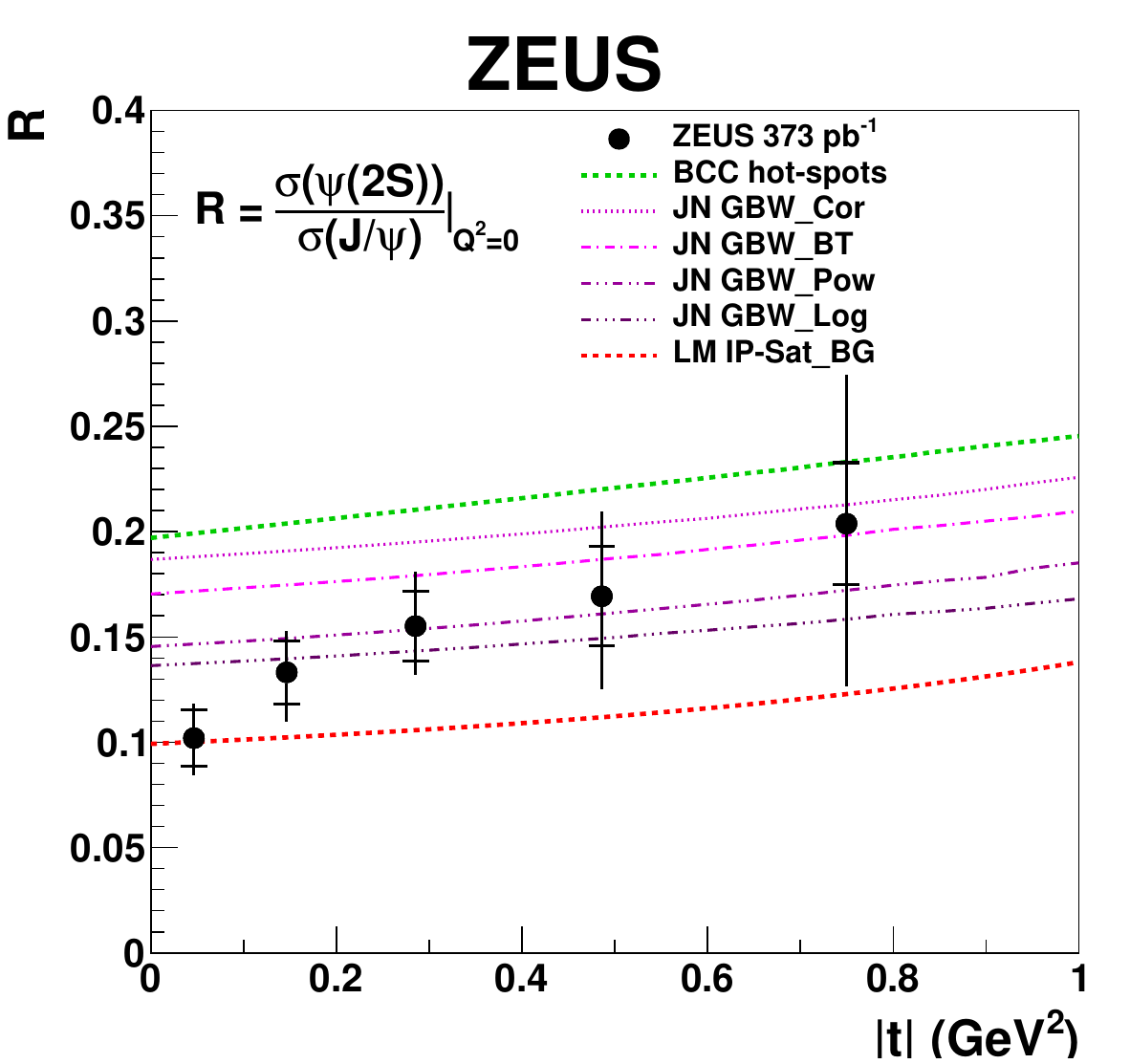}
\put(-40,220){\makebox(0,0)[tl]{(b)}}\\
\end{center}
 \caption{
Cross-section ratio $R = \sigma_{\psi(2S)}/\sigma_{J/\psi(1S)}$ in photoproduction for the combined $\psi(2S)$ decay modes as a 
function of (a) W and (b) |t|.  The ZEUS measurements are shown as solid circles.    The 
statistical uncertainties are shown as the inner error bars on the points whilst the outer error bars show the statistical and systematic uncertainties 
added in quadrature.  The points are shown at the mean $W$ and $|t|$ values for each bin as determined for the $J/\psi(1S)$ data (see Table~\ref{tab:tab1}). 
In (a) previous measurements from H1 (open 
points)~\protect\cite{pl:b541:251,pl:b421:385} are also shown.
Various QCD-inspired models are compared to the data and shown as lines (see Section~\ref{sect:Models} for details 
of the models).  No uncertainties for these predictions are provided.
}
\label{fig:R-W-t}
\vfill
\end{figure}

The cross-section ratios differential in $W$ and $|t|$ are shown in Fig.~\ref{fig:R-W-t} and given in Table~\ref{tab:tab1}.   
As a function of $W$, the value of $R$ is 
compatible with a constant value.  A slow increase 
of $R$ with increasing $|t|$ is observed.
The measurements presented in Fig.~\ref{fig:R-W-t}(a) are in agreement with previous measurements from H1~\cite{pl:b541:251,pl:b421:385}.  
In DIS~\cite{np:b909:934} neither a $W$ dependence nor a $|t|$ dependence of $R$ was observed.  A discussion of the comparison of the 
results to various model predictions is presented in Section~\ref{sect:Comparisons}.

\begin{figure}[p]
\vfill
\begin{center}
\includegraphics[width=16.0cm]{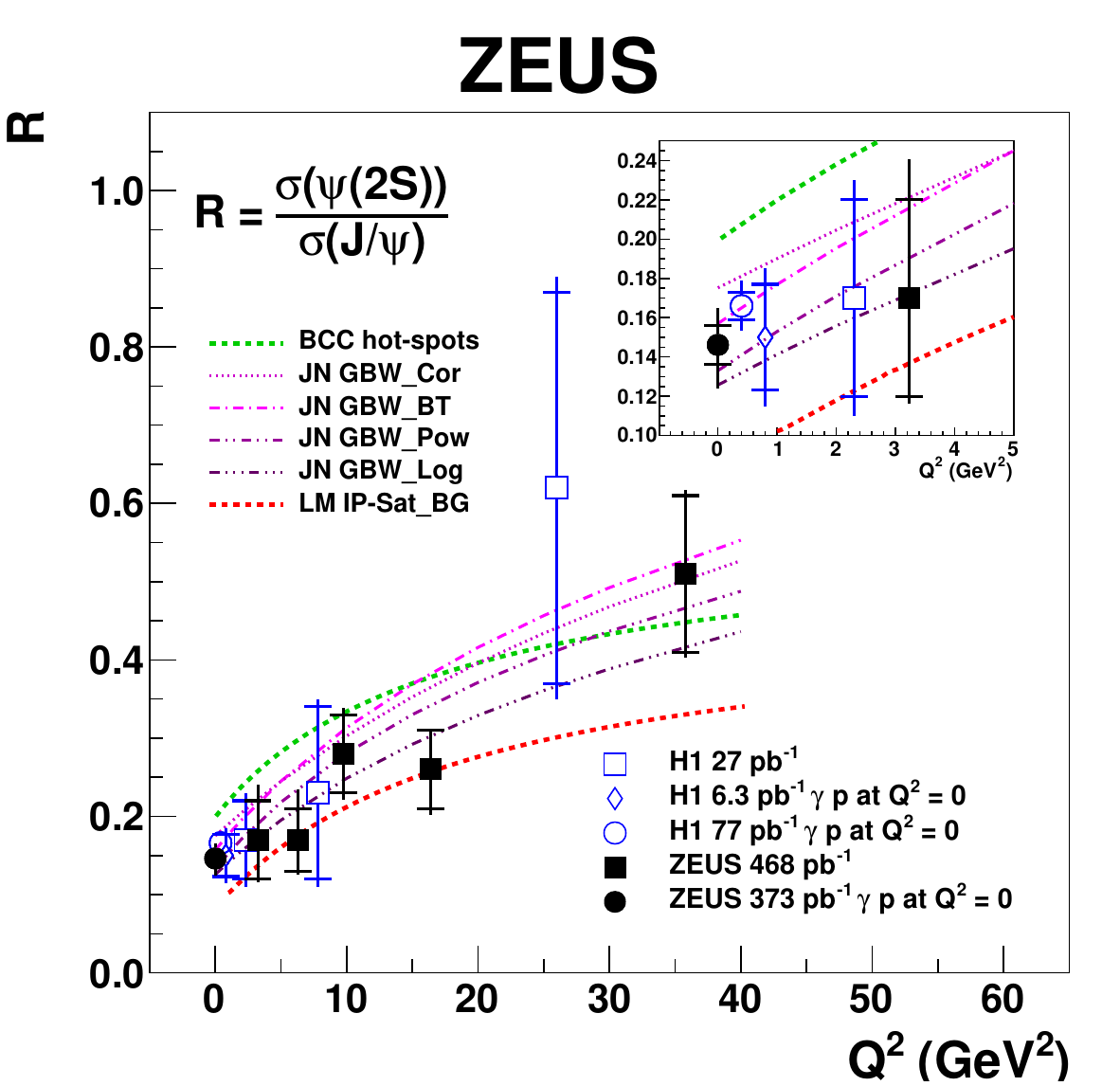}
\end{center}
 \caption{
Cross-section ratio $R = \sigma_{\psi(2S)}/\sigma_{J/\psi(1S)}$  as a function of $Q^2$.
The measurement from this analysis of photoproduction data for the combined $\psi(2S)$ decay modes is shown at $Q^2 = 0$\,GeV\,$^2$ 
(solid circle). Previous measurements are also shown in photoproduction from H1 (open circle and 
diamond)~\protect\cite{pl:b541:251,pl:b421:385} which are plotted 
horizontally displaced for better visualisation and measurements from both H1 (open squares)~\protect\cite{epj:c10:373} and ZEUS (solid  
squares)~\protect\cite{np:b909:934} in deep inelastic scattering.  The inset shows a zoom-in of the region at low $Q^2$ for better visibility.
The statistical uncertainties are shown as the inner error bars on the points whilst the outer error bars show the statistical and systematic uncertainties added in quadrature.  Various QCD-inspired models are compared to the data and shown as lines (see Section~\ref{sect:Models} for details 
of the models).  No uncertainties for these predictions are provided.
}
\label{fig:R-Q2}
\vfill
\end{figure}

The value of $R$ given above is shown in Fig.~\ref{fig:R-Q2} compared with other measurements in 
photoproduction and measurements in DIS as a function of $Q^2$.  The value measured here confirms the previous measurements in 
photoproduction~\cite{pl:b541:251,pl:b421:385}.  The trend of decreasing $R$ with decreasing $Q^2$ down to $\approx$\,0\,GeV$^2$ is also confirmed.

\section{Comparison to model predictions}
      \label{sect:Comparisons}

Several models of exclusive vector-meson production are available and also predict the ratio of the production of $\psi(2S)$ to $J/\psi(1S)$ 
mesons.  Predictions from three different models were compared to the data and are briefly described.  All models predictions were 
calculated for the kinematic region $30 < W < 180$\,GeV and $|t| < 1$\,GeV$^2$.

 \subsection{Individual models}
           \label{sect:Models}

The model from Bendov\'{a}, \v{C}epila and Contreras~\cite{pr:d99:034025} (BCC hot-spots) is based on energy-dependent hot spots, i.e.\ 
regions of high gluon density in the proton.  The slope parameter $b = 4.7$\,GeV$^{-2}$ was used for both $J/\psi(1S)$ and $\psi(2S)$ 
production. The $b$ value was derived from H1 and ZEUS data on $J/\psi(1S)$ photoproduction.

The model from Nemchik et al.~\cite{epj:c79:154,epj:c79:495,pr:d103:094027} (JN) provides predictions with various combinations 
of colour-dipole interactions, skewness parameters in the gluon density and quarkonia potentials used for the calculation of the 
centre-of-mass wave functions.  The predictions shown are based on the Golec-Biernat--W\"{u}sthoff (GBW) colour-dipole 
model~\cite{pr:d59:014017,pr:d60:114023} 
with skewness.  The phenomenological quarkonia potentials used were: the so-called Buchm\"{u}ller--Tye (BT), logarithmic (Log), 
Cornell (Cor) and power-law (Pow).  Other combinations of colour-dipole models with or without skewness differ to those shown 
by 5--10\%.

Lappi and M\"{a}ntysaari~\cite{pr:c83:065202} (LM) use the BFKL evolution as well as the IP-Sat 
model~\cite{pr:d87:034002} to predict vector-meson production in $ep$ and electron--ion collisions in the dipole picture. 
The wave functions of the $J/\psi(1S)$ and $\psi(2S)$ have been calculated according to the boosted Gaussian (BG) 
procedure~\cite{Kowalski:2006hc,pos:dis2014:069} and the low-$x$ inclusive HERA data have been used to constrain 
the $c\bar{c}$--dipole cross section.

 \subsection{Comparison of models and data}
           \label{sect:Data_comparison}

In Fig.~\ref{fig:R-W-t}, model predictions are compared to photoproduction data as a function of $W$ and $|t|$.  
All model predictions exhibit a mild rise in $R$ with increasing $W$.  The predicted rise is similar for all models.  The 
absolute values of the predictions differ by up to a factor of 2.   The predictions from BCC lie above 
the data and the predictions from LM lie below the data.  No uncertainties for these predictions are provided.  
The shapes of the models are consistent with the data, although the data are also consistent with no increase with $W$.  
The predictions from JN give a better description of the normalisation and the differences in predictions due 
to the quarkonia potential also give some indication of the uncertainty in the models.

All models also predict an increase in $R$ with increasing $|t|$, and again predict similar gradients but different absolute 
values.  Given the uncertainties in the data and the spread of the models, the description of the data is good.

In Fig.~\ref{fig:R-Q2}, model predictions are compared to photoproduction and DIS data as a function of $Q^2$.  
All models predict a strong 
increase in $R$ with increasing $Q^2$, which is compatible with the trend seen in the data.  
Towards higher $Q^2$, the LM and BCC models exhibit a flattening of $R$ compared to the JN models.  
The photoproduction data have the potential to constrain the models further.

Overall, the predictions from the three models, BCC, JN and LM, give a reasonable description of the $W$, $|t|$ and $Q^2$ 
dependence of $R$.

 \section{Summary}
      \label{sect:Summary}
The cross-section ratio $R = \sigma _{\psi(2S)}/ \sigma _{J/\psi(1S)}$ in exclusive photoproduction has been measured with the ZEUS 
detector at HERA in the kinematic range $Q^2 <1$\,GeV$^2$, $30 <W <180$\,GeV and $|t| < 1$\,GeV$^2$, using an integrated 
luminosity of 373\,pb$^{-1}$.  The decay channels used were $\mu^+ \mu^-$ and $ J/\psi(1S) \,\pi^+ \pi^-$ for the $\psi(2S)$ and
$\mu^+ \mu^-$ for the $ J/\psi (1S)$.  The cross-section ratio was determined as a function of $W$ and $|t|$ and presented as a function 
of $Q^2$.  As a function 
of $W$, the value of $R$ is compatible with a constant value.  A slow increase 
of $R$ with increasing $|t|$ is observed.  The data confirm previous conclusions that $R$ decreases with 
decreasing $Q^2$.  Three model calculations were compared to the measured dependences of $R$ and give a reasonable description 
of the data, which can be used to constrain the models further.

\section*{Acknowledgements}
\label{sec-ack}

We appreciate the contributions to the construction, maintenance and operation of the ZEUS detector of many people who are not listed as 
authors. The HERA machine group and the DESY computing staff are especially acknowledged for their success in providing excellent 
operation of the collider and the data-analysis environment.  We thank the DESY directorate for their strong support and encouragement.
We thank Dagmar Bendov\'{a}, Jan \v{C}epila, Michal K\v{r}elina, Jan Nemchik and Heikki M\"{a}ntysaari for providing model 
predictions and for useful discussions. 

\clearpage

{
\ifzeusbst
  \ifzmcite
     \bibliographystyle{./BiBTeX/bst/l4z_default3}
  \else
     \bibliographystyle{./BiBTeX/bst/l4z_default3_nomcite}
  \fi
\fi
\ifzdrftbst
  \ifzmcite
    \bibliographystyle{./BiBTeX/bst/l4z_draft3}
  \else
    \bibliographystyle{./BiBTeX/bst/l4z_draft3_nomcite}
  \fi
\fi
\ifzbstepj
  \ifzmcite
    \bibliographystyle{./BiBTeX/bst/l4z_epj3}
  \else
    \bibliographystyle{./BiBTeX/bst/l4z_epj3_nomcite}
  \fi
\fi
\ifzbstjhep
  \ifzmcite
    \bibliographystyle{./BiBTeX/bst/l4z_jhep3}
  \else
    \bibliographystyle{./BiBTeX/bst/l4z_jhep3_nomcite}
  \fi
\fi
\ifzbstnp
  \ifzmcite
    \bibliographystyle{./BiBTeX/bst/l4z_np3}
  \else
    \bibliographystyle{./BiBTeX/bst/l4z_np3_nomcite}
  \fi
\fi
\ifzbstpl
  \ifzmcite
    \bibliographystyle{./BiBTeX/bst/l4z_pl3}
  \else
    \bibliographystyle{./BiBTeX/bst/l4z_pl3_nomcite}
  \fi
\fi
{\raggedright
\bibliography{./BiBTeX/bib/l4z_zeus.bib,%
              ./BiBTeX/bib/l4z_h1.bib,%
              ./BiBTeX/bib/l4z_articles.bib,%
              ./BiBTeX/bib/l4z_books.bib,%
              ./BiBTeX/bib/l4z_conferences.bib,%
              ./BiBTeX/bib/l4z_misc.bib,%
              ./BiBTeX/bib/l4z_preprints.bib}}
}
\vfill\eject

\setcounter{table}{0}
\setlength{\topmargin}{-4.5cm}
\newgeometry{textheight=25cm,textwidth=448.13095pt}

\newcommand\Tstrut{\rule{0pt}{2.2ex}}         
\newcommand\Bstrut{\rule[-0.9ex]{0pt}{0pt}}   



%
%
\begin{table} \centering \tiny
\begin{sideways}
\begin{tabular}{|c|c|c|c|c|c|c|c|c|c||c|} \hline
bin & channel & $\langle W \rangle , \langle t \rangle$ & signal & bg-nonres & bg-res & $A$ & $f_{\rm pdiss}$ & $\langle f_{\rm pdiss}^{\psi(2S)} \rangle$ & $R_{\mu\mu},\,R_{J/\psi\,\pi\pi}$ & $R$ \Tstrut\\ \hline\hline
All & $J/\psi(1S)\rightarrow \mu^+\mu^-$ & $ 91.4,\, 0.219$ & $23043 \pm 169$ & $2106 \pm 46$ & $550 \pm 40$ & $0.090 \pm 0.0003$ & $0.168 \pm 0.004$ & & & $0.146$\Tstrut\\ 
events & $\psi(2S)\rightarrow \mu^+\mu^-$ & $ 95.8,\, 0.223$ & $690 \pm 50$ & $1701 \pm 41$ & $103 \pm 10$ & $0.132 \pm 0.0008$ & $0.15 \pm 0.02$ & $0.16 \pm 0.01$ & $0.154 \pm 0.012$ & $ \pm 0.010\, \mathrm{(stat.)}$ \Tstrut\\ 
 & $\psi(2S)\rightarrow \mu^+\mu^-\pi^+\pi^-$ & $ 99.6,\, 0.225$ & $387 \pm 20$ & $10 \pm 3$ & $ $ & $0.035 \pm 0.0050$ & $0.18 \pm 0.02$ & $ $ & $0.125 \pm 0.019$ & $ ^{+0.016}_{-0.020}\, \mathrm{(syst.)}$ \Tstrut\\ \hline\hline
W1 & $J/\psi(1S)\rightarrow \mu^+\mu^-$ & $ 45.5$ & $7301 \pm 104$ & $677 \pm 26$ & $148 \pm 27$ & $0.087 \pm 0.0006$ & $0.168 \pm 0.004$ & & & $0.128$\Tstrut\\ 
(30,60) & $\psi(2S)\rightarrow \mu^+\mu^-$ & $ 48.1$ & $178 \pm 33$ & $500 \pm 22$ & $38 \pm 6$ & $0.115 \pm 0.0014$ & $0.15 \pm 0.02$ & $0.16 \pm 0.01$ & $0.138 \pm 0.025$ & $ \pm 0.017\, \mathrm{(stat.)}$ \Tstrut\\ 
GeV & $\psi(2S)\rightarrow \mu^+\mu^-\pi^+\pi^-$ & $ 50.7$ & $68 \pm 8$ & $1.0 \pm 1.0$ & $ $ & $0.020 \pm 0.0030$ & $0.18 \pm 0.02$ & $ $ & $0.120 \pm 0.024$ & $ ^{+0.013}_{-0.025}\, \mathrm{(syst.)}$ \Tstrut\\ \hline
W2 & $J/\psi(1S)\rightarrow \mu^+\mu^-$ & $ 74.6$ & $3985 \pm 80$ & $342 \pm 18$ & $136 \pm 17$ & $0.072 \pm 0.0006$ & $0.168 \pm 0.004$ & & & $0.123$\Tstrut\\ 
(60,90) & $\psi(2S)\rightarrow \mu^+\mu^-$ & $ 73.9$ & $123 \pm 16$ & $404 \pm 20$ & $1 \pm 1$ & $0.121 \pm 0.0016$ & $0.15 \pm 0.02$ & $0.16 \pm 0.01$ & $0.137 \pm 0.018$ & $ \pm 0.013\, \mathrm{(stat.)}$ \Tstrut\\ 
GeV & $\psi(2S)\rightarrow \mu^+\mu^-\pi^+\pi^-$ & $ 75.2$ & $87 \pm 10$ & $3.8 \pm 2.0$ & $ $ & $0.043 \pm 0.0065$ & $0.18 \pm 0.02$ & $ $ & $0.105 \pm 0.020$ & $ ^{+0.015}_{-0.016}\, \mathrm{(syst.)}$ \Tstrut\\ \hline
W3 & $J/\psi(1S)\rightarrow \mu^+\mu^-$ & $ 105.6$ & $5278 \pm 81$ & $561 \pm 24$ & $164 \pm 19$ & $0.115 \pm 0.0008$ & $0.168 \pm 0.004$ & & & $0.164$\Tstrut\\ 
(90,120) & $\psi(2S)\rightarrow \mu^+\mu^-$ & $ 105.3$ & $183 \pm 21$ & $401 \pm 20$ & $3 \pm 2$ & $0.166 \pm 0.0020$ & $0.15 \pm 0.02$ & $0.16 \pm 0.01$ & $0.180 \pm 0.021$ & $ \pm 0.016\, \mathrm{(stat.)}$ \Tstrut\\ 
GeV & $\psi(2S)\rightarrow \mu^+\mu^-\pi^+\pi^-$ & $ 105.6$ & $114 \pm 11$ & $2.5 \pm 1.6$ & $ $ & $0.051 \pm 0.0080$ & $0.18 \pm 0.02$ & $ $ & $0.140 \pm 0.026$ & $ ^{+0.019}_{-0.022}\, \mathrm{(syst.)}$ \Tstrut\\ \hline
W4 & $J/\psi(1S)\rightarrow \mu^+\mu^-$ & $ 129.7$ & $3333 \pm 65$ & $294 \pm 17$ & $60 \pm 12$ & $0.127 \pm 0.0011$ & $0.168 \pm 0.004$ & & & $0.123$\Tstrut\\ 
(120,140) & $\psi(2S)\rightarrow \mu^+\mu^-$ & $ 129.4$ & $70 \pm 14$ & $198 \pm 14$ & $11 \pm 3$ & $0.162 \pm 0.0026$ & $0.15 \pm 0.02$ & $0.16 \pm 0.01$ & $0.124 \pm 0.025$ & $ \pm 0.018\, \mathrm{(stat.)}$ \Tstrut\\ 
GeV & $\psi(2S)\rightarrow \mu^+\mu^-\pi^+\pi^-$ & $ 129.3$ & $64 \pm 8$ & $1.6 \pm 1.3$ & $ $ & $0.059 \pm 0.0105$ & $0.18 \pm 0.02$ & $ $ & $0.121 \pm 0.027$ & $ ^{+0.016}_{-0.018}\, \mathrm{(syst.)}$ \Tstrut\\ \hline
W5 & $J/\psi(1S)\rightarrow \mu^+\mu^-$ & $ 153.1$ & $3261 \pm 71$ & $202 \pm 14$ & $49 \pm 10$ & $0.074 \pm 0.0006$ & $0.168 \pm 0.004$ & & & $0.174$\Tstrut\\ 
(140,180) & $\psi(2S)\rightarrow \mu^+\mu^-$ & $ 155.4$ & $127 \pm 27$ & $178 \pm 13$ & $54 \pm 7$ & $0.112 \pm 0.0017$ & $0.15 \pm 0.02$ & $0.16 \pm 0.01$ & $0.191 \pm 0.041$ & $ \pm 0.028\, \mathrm{(stat.)}$ \Tstrut\\ 
GeV & $\psi(2S)\rightarrow \mu^+\mu^-\pi^+\pi^-$ & $ 153.2$ & $54 \pm 7$ & $1.0 \pm 1.0$ & $ $ & $0.022 \pm 0.0043$ & $0.18 \pm 0.02$ & $ $ & $0.159 \pm 0.038$ & $ ^{+0.021}_{-0.021}\, \mathrm{(syst.)}$ \Tstrut\\ \hline
\hline
t1 & $J/\psi(1S)\rightarrow \mu^+\mu^-$ & $ 0.046$ & $7949 \pm 107$ & $1481 \pm 38$ & $165 \pm 30$ & $0.092 \pm 0.0005$ & $0.079 \pm 0.002$ & & & $0.102$\Tstrut\\ 
(0.0,0.1) & $\psi(2S)\rightarrow \mu^+\mu^-$ & $ 0.047$ & $176 \pm 33$ & $1238 \pm 35$ & $25 \pm 5$ & $0.132 \pm 0.0014$ & $0.06 \pm 0.01$ & $0.07 \pm 0.01$ & $0.115 \pm 0.021$ & $ \pm 0.013\, \mathrm{(stat.)}$ \Tstrut\\ 
GeV$^{2}$ & $\psi(2S)\rightarrow \mu^+\mu^-\pi^+\pi^-$ & $ 0.046$ & $108 \pm 11$ & $2.8 \pm 1.7$ & $ $ & $0.039 \pm 0.0058$ & $0.08 \pm 0.01$ & $ $ & $0.094 \pm 0.017$ & $ ^{+0.010}_{-0.011}\, \mathrm{(syst.)}$ \Tstrut\\ \hline
t2 & $J/\psi(1S)\rightarrow \mu^+\mu^-$ & $ 0.146$ & $4997 \pm 77$ & $154 \pm 12$ & $109 \pm 15$ & $0.092 \pm 0.0006$ & $0.105 \pm 0.002$ & & & $0.133$\Tstrut\\ 
(0.1,0.2) & $\psi(2S)\rightarrow \mu^+\mu^-$ & $ 0.147$ & $134 \pm 18$ & $152 \pm 12$ & $16 \pm 4$ & $0.135 \pm 0.0017$ & $0.09 \pm 0.01$ & $0.10 \pm 0.01$ & $0.138 \pm 0.019$ & $ \pm 0.015\, \mathrm{(stat.)}$ \Tstrut\\ 
GeV$^{2}$ & $\psi(2S)\rightarrow \mu^+\mu^-\pi^+\pi^-$ & $ 0.147$ & $82 \pm 9$ & $2.4 \pm 1.5$ & $ $ & $0.035 \pm 0.0056$ & $0.11 \pm 0.01$ & $ $ & $0.125 \pm 0.024$ & $ ^{+0.013}_{-0.018}\, \mathrm{(syst.)}$ \Tstrut\\ \hline
t3 & $J/\psi(1S)\rightarrow \mu^+\mu^-$ & $ 0.285$ & $5239 \pm 81$ & $199 \pm 14$ & $143 \pm 19$ & $0.090 \pm 0.0006$ & $0.164 \pm 0.003$ & & & $0.155$\Tstrut\\ 
(0.2,0.4) & $\psi(2S)\rightarrow \mu^+\mu^-$ & $ 0.286$ & $159 \pm 21$ & $190 \pm 14$ & $24 \pm 5$ & $0.133 \pm 0.0016$ & $0.13 \pm 0.02$ & $0.15 \pm 0.01$ & $0.156 \pm 0.021$ & $ \pm 0.017\, \mathrm{(stat.)}$ \Tstrut\\ 
GeV$^{2}$ & $\psi(2S)\rightarrow \mu^+\mu^-\pi^+\pi^-$ & $ 0.284$ & $100 \pm 10$ & $2.7 \pm 1.6$ & $ $ & $0.033 \pm 0.0051$ & $0.18 \pm 0.02$ & $ $ & $0.154 \pm 0.029$ & $ ^{+0.020}_{-0.016}\, \mathrm{(syst.)}$ \Tstrut\\ \hline
t4 & $J/\psi(1S)\rightarrow \mu^+\mu^-$ & $ 0.486$ & $2588 \pm 54$ & $60 \pm 8$ & $62 \pm 10$ & $0.088 \pm 0.0009$ & $0.275 \pm 0.006$ & & & $0.169$\Tstrut\\ 
(0.4,0.6) & $\psi(2S)\rightarrow \mu^+\mu^-$ & $ 0.486$ & $102 \pm 17$ & $66 \pm 8$ & $15 \pm 4$ & $0.128 \pm 0.0023$ & $0.23 \pm 0.03$ & $0.26 \pm 0.02$ & $0.207 \pm 0.035$ & $ \pm 0.024\, \mathrm{(stat.)}$ \Tstrut\\ 
GeV$^{2}$ & $\psi(2S)\rightarrow \mu^+\mu^-\pi^+\pi^-$ & $ 0.486$ & $45 \pm 7$ & $0.5 \pm 0.7$ & $ $ & $0.032 \pm 0.0056$ & $0.29 \pm 0.03$ & $ $ & $0.138 \pm 0.032$ & $ ^{+0.032}_{-0.038}\, \mathrm{(syst.)}$ \Tstrut\\ \hline
t5 & $J/\psi(1S)\rightarrow \mu^+\mu^-$ & $ 0.750$ & $2459 \pm 55$ & $73 \pm 9$ & $76 \pm 11$ & $0.080 \pm 0.0011$ & $0.469 \pm 0.010$ & & & $0.204$\Tstrut\\ 
(0.6,1.0) & $\psi(2S)\rightarrow \mu^+\mu^-$ & $ 0.748$ & $121 \pm 18$ & $77 \pm 9$ & $23 \pm 5$ & $0.119 \pm 0.0028$ & $0.43 \pm 0.06$ & $0.45 \pm 0.04$ & $0.254 \pm 0.043$ & $ \pm 0.029\, \mathrm{(stat.)}$ \Tstrut\\ 
GeV$^{2}$ & $\psi(2S)\rightarrow \mu^+\mu^-\pi^+\pi^-$ & $ 0.755$ & $53 \pm 7$ & $1.5 \pm 1.2$ & $ $ & $0.031 \pm 0.0055$ & $0.47 \pm 0.06$ & $ $ & $0.163 \pm 0.039$ & $ ^{+0.070}_{-0.071}\, \mathrm{(syst.)}$ \Tstrut\\ \hline
\end{tabular}
\end{sideways}
\caption{Table of results with columns showing the $W$ and $|t|$ bins, the decay channel, the mean values, $\langle W \rangle$ and 
$\langle t \rangle$, the number of signal events, the number of background events from Bethe--Heitler events and combinatorial 
non-resonant background (bg-nonres) and resonant background, described in Section~\ref{sec:signal} (bg-res), the acceptance ($A$), the fraction of proton-dissociative events ($f_{\rm pdiss}$), 
the average fraction of events 
from proton dissociation for the two $\psi(2S)$ decay channels ($\langle f_{\rm pdiss}^{\psi(2S)} \rangle$), the cross-section ratios for 
$R_{\mu \mu}$ for $\psi (2S ) \rightarrow \mu ^+ \mu^-$,  $R_{J/\psi \,\pi \pi} $ for $\psi (2S ) \rightarrow J/\psi (1S ) \,\pi^+ \pi^-$ and 
$R$ for the combination of the two decay modes.  Statistical and systematic uncertainties are given separately for $R$; all other 
uncertainties are statistical only.}
\label{tab:tab1}
\end{table}
%



\setlength{\topmargin}{0cm}
\restoregeometry
\setlength{\topmargin}{-2.0cm}




\newcommand\TStrut{\rule{0pt}{3.2ex}}         
\newcommand\BStrut{\rule[-1.6ex]{0pt}{0pt}}   


%
%
\begin{table} \centering \scriptsize
\begin{sideways}
\begin{tabular}{|l||c||c|c|c|c|c||c|c|c|c|c|} \hline
variation & $R$ & $R|_{W1}$ & $R|_{W2}$ & $R|_{W3}$ & $R|_{W4}$ & $R|_{W5}$ & $R|_{t1}$ & $R|_{t2}$ & $R|_{t3}$ & $R|_{t4}$ & $R|_{t5}$ \TStrut\BStrut\\ \hline\hline
$\Delta_{1}:$ $b_{el}^{J/\psi}\pm 0.3\,\mathrm{GeV}^{-2}$ & $^{+0.0022}_{-0.0023}$ & $^{+0.0019}_{-0.0019}$ & $^{+0.0017}_{-0.0017}$ & $^{+0.0030}_{-0.0031}$ & $^{+0.0019}_{-0.0020}$ & $^{+0.0021}_{-0.0022}$ & $^{+0.0003}_{-0.0003}$ & $^{+0.0009}_{-0.0009}$ & $^{+0.0027}_{-0.0026}$ & $^{+0.0075}_{-0.0067}$ & $^{+0.0226}_{-0.0191}$ \TStrut\BStrut\\ \hline
$\Delta_{2}:$ $b_{el}^{\psi(2S)}\pm 0.7\,\mathrm{GeV}^{-2}$ & $^{-0.0041}_{+0.0055}$ & $^{-0.0033}_{+0.0046}$ & $^{-0.0030}_{+0.0040}$ & $^{-0.0055}_{+0.0075}$ & $^{-0.0035}_{+0.0047}$ & $^{-0.0039}_{+0.0054}$ & $^{-0.0002}_{+0.0006}$ & $^{-0.0014}_{+0.0020}$ & $^{-0.0043}_{+0.0051}$ & $^{-0.0145}_{+0.0152}$ & $^{-0.0422}_{+0.0466}$ \TStrut\BStrut\\ \hline
$\Delta_{3}:$ $b_{pd}^{J/\psi}\pm 0.1\,\mathrm{GeV}^{-2}$ & $^{+0.0042}_{-0.0038}$ & $^{+0.0037}_{-0.0034}$ & $^{+0.0036}_{-0.0032}$ & $^{+0.0048}_{-0.0043}$ & $^{+0.0036}_{-0.0032}$ & $^{+0.0050}_{-0.0046}$ & $^{+0.0017}_{-0.0015}$ & $^{+0.0028}_{-0.0025}$ & $^{+0.0047}_{-0.0042}$ & $^{+0.0076}_{-0.0070}$ & $^{+0.0126}_{-0.0121}$ \TStrut\BStrut\\ \hline
$\Delta_{4}:$ $b_{pd}^{\psi(2S)}\pm 0.2\,\mathrm{GeV}^{-2}$ & $^{-0.0117}_{+0.0106}$ & $^{-0.0101}_{+0.0092}$ & $^{-0.0098}_{+0.0088}$ & $^{-0.0130}_{+0.0118}$ & $^{-0.0097}_{+0.0088}$ & $^{-0.0137}_{+0.0124}$ & $^{-0.0044}_{+0.0034}$ & $^{-0.0075}_{+0.0060}$ & $^{-0.0123}_{+0.0108}$ & $^{-0.0197}_{+0.0199}$ & $^{-0.0342}_{+0.0419}$ \TStrut\BStrut\\ \hline
$\Delta_{5}:$ $\alpha' \pm 0.04\,\mathrm{GeV}^{-2}$ & $^{+0.0002}_{+0.0005}$ & $^{+0.0002}_{+0.0004}$ & $^{+0.0002}_{+0.0004}$ & $^{+0.0003}_{+0.0005}$ & $^{+0.0002}_{+0.0004}$ & $^{+0.0002}_{+0.0006}$ & $^{+0.0001}_{+0.0002}$ & $^{+0.0001}_{+0.0002}$ & $^{+0.0002}_{+0.0005}$ & $^{+0.0004}_{+0.0011}$ & $^{+0.0003}_{+0.0020}$ \TStrut\BStrut\\ \hline
$\Delta_{6}:$ $\delta_{el}^{J/\psi}\pm 0.1$ & $^{+0.0005}_{-0.0006}$ & $^{+0.0005}_{-0.0006}$ & $^{+0.0001}_{-0.0002}$ & $^{+0.0002}_{-0.0002}$ & $^{+0.0000}_{-0.0001}$ & $^{-0.0006}_{+0.0006}$ & $^{+0.0003}_{-0.0004}$ & $^{+0.0005}_{-0.0005}$ & $^{+0.0006}_{-0.0007}$ & $^{+0.0008}_{-0.0008}$ & $^{+0.0016}_{-0.0017}$ \TStrut\BStrut\\ \hline
$\Delta_{7}:$ $\delta_{el}^{\psi(2S)}\pm 0.2$ & $^{-0.0007}_{+0.0019}$ & $^{-0.0026}_{+0.0035}$ & $^{+0.0005}_{+0.0002}$ & $^{+0.0002}_{+0.0007}$ & $^{+0.0004}_{+0.0002}$ & $^{+0.0015}_{-0.0006}$ & $^{-0.0006}_{+0.0012}$ & $^{-0.0004}_{+0.0012}$ & $^{-0.0009}_{+0.0023}$ & $^{-0.0007}_{+0.0030}$ & $^{-0.0026}_{+0.0061}$ \TStrut\BStrut\\ \hline
$\Delta_{8}:$ $\delta_{pd}^{J/\psi}\pm 0.15$ & $^{-0.0000}_{+0.0000}$ & $^{-0.0000}_{+0.0000}$ & $^{-0.0000}_{+0.0000}$ & $^{-0.0000}_{+0.0000}$ & $^{-0.0000}_{+0.0000}$ & $^{-0.0000}_{+0.0000}$ & $^{-0.0000}_{+0.0000}$ & $^{-0.0000}_{+0.0000}$ & $^{-0.0000}_{+0.0001}$ & $^{-0.0000}_{+0.0000}$ & $^{-0.0001}_{+0.0001}$ \TStrut\BStrut\\ \hline
$\Delta_{9}:$ $\delta_{pd}^{\psi(2S)}\pm 0.3$ & $^{+0.0004}_{+0.0004}$ & $^{+0.0003}_{+0.0003}$ & $^{+0.0003}_{+0.0003}$ & $^{+0.0004}_{+0.0004}$ & $^{+0.0003}_{+0.0003}$ & $^{+0.0004}_{+0.0005}$ & $^{+0.0002}_{+0.0001}$ & $^{+0.0002}_{+0.0002}$ & $^{+0.0005}_{+0.0004}$ & $^{+0.0006}_{+0.0011}$ & $^{+0.0009}_{+0.0016}$ \TStrut\BStrut\\ \hline
$\Delta_{10}:$ $\beta^{J/\psi}\pm 0.3$ & $^{-0.0001}_{+0.0001}$ & $^{-0.0001}_{+0.0001}$ & $^{-0.0001}_{+0.0001}$ & $^{-0.0001}_{+0.0001}$ & $^{-0.0001}_{+0.0001}$ & $^{-0.0002}_{+0.0002}$ & $^{-0.0001}_{+0.0001}$ & $^{-0.0001}_{+0.0001}$ & $^{-0.0001}_{+0.0001}$ & $^{-0.0002}_{+0.0002}$ & $^{-0.0004}_{+0.0004}$ \TStrut\BStrut\\ \hline
$\Delta_{11}:$ $\beta^{\psi(2S)}\pm 0.3$ & $^{+0.0006}_{+0.0001}$ & $^{+0.0005}_{+0.0001}$ & $^{+0.0005}_{+0.0001}$ & $^{+0.0006}_{+0.0001}$ & $^{+0.0005}_{+0.0001}$ & $^{+0.0007}_{+0.0001}$ & $^{+0.0002}_{+0.0001}$ & $^{+0.0004}_{-0.0000}$ & $^{+0.0006}_{+0.0002}$ & $^{+0.0012}_{+0.0003}$ & $^{+0.0020}_{+0.0002}$ \TStrut\BStrut\\ \hline
$\Delta_{12}:$ $(p_Z^{\mu}/p_T^{\mu}/p^{\mu}, \eta^{\mu})$ grid & $^{-0.0001}_{-----}$ & $^{-0.0015}_{-----}$ & $^{-0.0007}_{-----}$ & $^{+0.0010}_{-----}$ & $^{+0.0001}_{-----}$ & $^{+0.0015}_{-----}$ & $^{+0.0004}_{-----}$ & $^{-0.0003}_{-----}$ & $^{+0.0001}_{-----}$ & $^{-0.0018}_{-----}$ & $^{+0.0006}_{-----}$ \TStrut\BStrut\\ \hline
$\Delta_{13}:$ $p_T^{\mu}\pm 0.1\, \mathrm{GeV}$ & $^{+0.0024}_{-0.0002}$ & $^{-0.0054}_{-0.0044}$ & $^{-0.0048}_{+0.0054}$ & $^{+0.0010}_{+0.0015}$ & $^{-0.0010}_{-0.0003}$ & $^{-0.0054}_{+0.0075}$ & $^{+0.0048}_{+0.0030}$ & $^{-0.0066}_{+0.0073}$ & $^{-0.0014}_{+0.0098}$ & $^{-0.0078}_{+0.0091}$ & $^{-0.0102}_{+0.0029}$ \TStrut\BStrut\\ \hline
$\Delta_{14}:$ $p_T^{\pi}\pm 0.02\, \mathrm{GeV}$ & $^{-0.0012}_{+0.0034}$ & $^{-0.0020}_{+0.0005}$ & $^{+0.0007}_{+0.0060}$ & $^{-0.0029}_{+0.0023}$ & $^{-0.0013}_{+0.0048}$ & $^{-0.0034}_{+0.0089}$ & $^{-0.0026}_{+0.0039}$ & $^{-0.0003}_{+0.0025}$ & $^{-0.0006}_{+0.0025}$ & $^{-0.0046}_{+0.0121}$ & $^{-0.0024}_{+0.0052}$ \TStrut\BStrut\\ \hline
$\Delta_{15}:$ $\pi^{\pm}$ vertex & $^{-0.0065}_{-----}$ & $^{-0.0127}_{-----}$ & $^{-0.0072}_{-----}$ & $^{-0.0063}_{-----}$ & $^{-0.0107}_{-----}$ & $^{-0.0108}_{-----}$ & $^{-0.0062}_{-----}$ & $^{-0.0112}_{-----}$ & $^{-0.0034}_{-----}$ & $^{-0.0127}_{-----}$ & $^{-0.0330}_{-----}$ \TStrut\BStrut\\ \hline
$\Delta_{16}:$ $\pi^{\pm}$ $p_T$ correction $\pm 0.04\,\mathrm{GeV}$ & $^{-0.0013}_{+0.0021}$ & $^{-0.0019}_{+0.0029}$ & $^{-0.0019}_{+0.0028}$ & $^{-0.0021}_{+0.0033}$ & $^{-0.0015}_{+0.0024}$ & $^{-0.0025}_{+0.0039}$ & $^{-0.0018}_{+0.0027}$ & $^{-0.0014}_{+0.0021}$ & $^{-0.0015}_{+0.0024}$ & $^{-0.0029}_{+0.0049}$ & $^{-0.0037}_{+0.0063}$ \TStrut\BStrut\\ \hline
$\Delta_{17}:$ elasticity threshold $\pm 0.1\,\mathrm{GeV}$ & $^{+0.0016}_{-0.0040}$ & $^{+0.0022}_{-0.0042}$ & $^{+0.0029}_{-0.0053}$ & $^{+0.0012}_{-0.0043}$ & $^{+0.0064}_{-0.0069}$ & $^{+0.0039}_{+0.0005}$ & $^{+0.0034}_{-0.0066}$ & $^{+0.0013}_{-0.0028}$ & $^{+0.0014}_{+0.0014}$ & $^{+0.0004}_{-0.0084}$ & $^{+0.0111}_{-0.0044}$ \TStrut\BStrut\\ \hline
$\Delta_{18}:$ forward cone $\theta \pm 0.02\,\mathrm{rad}$ & $^{+0.0006}_{+0.0006}$ & $^{+0.0010}_{+0.0005}$ & $^{-0.0002}_{+0.0003}$ & $^{+0.0009}_{+0.0005}$ & $^{+0.0002}_{+0.0003}$ & $^{+0.0003}_{+0.0004}$ & $^{+0.0000}_{+0.0000}$ & $^{+0.0007}_{+0.0002}$ & $^{+0.0002}_{+0.0007}$ & $^{+0.0027}_{+0.0007}$ & $^{+0.0006}_{+0.0017}$ \TStrut\BStrut\\ \hline
$\Delta_{19}:$ forward energy $\pm 0.25\,\mathrm{GeV}$ & $^{+0.0015}_{-0.0001}$ & $^{+0.0017}_{-0.0008}$ & $^{-0.0000}_{-0.0006}$ & $^{+0.0007}_{+0.0013}$ & $^{+0.0014}_{+0.0004}$ & $^{+0.0009}_{+0.0026}$ & $^{+0.0006}_{-0.0012}$ & $^{-0.0006}_{+0.0012}$ & $^{+0.0030}_{+0.0011}$ & $^{-0.0002}_{+0.0028}$ & $^{+0.0054}_{-0.0021}$ \TStrut\BStrut\\ \hline
$\Delta_{20}:$ muons timing $\pm 1.0\, \mathrm{ns}$ & $^{-0.0002}_{+0.0009}$ & $^{+0.0000}_{+0.0004}$ & $^{-0.0008}_{+0.0009}$ & $^{+0.0015}_{-0.0003}$ & $^{+0.0007}_{+0.0002}$ & $^{+0.0004}_{-0.0002}$ & $^{-0.0006}_{+0.0009}$ & $^{+0.0017}_{-0.0014}$ & $^{+0.0011}_{+0.0018}$ & $^{+0.0020}_{-0.0010}$ & $^{+0.0041}_{+0.0014}$ \TStrut\BStrut\\ \hline
$\Delta_{21}:$ MC tempates fit & $^{-0.0097}_{-----}$ & $^{-0.0134}_{-----}$ & $^{-0.0008}_{-----}$ & $^{-0.0123}_{-----}$ & $^{+0.0064}_{-----}$ & $^{-0.0037}_{-----}$ & $^{-0.0005}_{-----}$ & $^{-0.0067}_{-----}$ & $^{+0.0038}_{-----}$ & $^{-0.0179}_{-----}$ & $^{-0.0185}_{-----}$ \TStrut\BStrut\\ \hline
$\Delta_{22}:$ $B_{J/\psi \rightarrow \mu^+\mu^-} \pm 1\sigma$ & $^{+0.0008}_{-0.0008}$ & $^{+0.0007}_{-0.0007}$ & $^{+0.0007}_{-0.0007}$ & $^{+0.0009}_{-0.0009}$ & $^{+0.0007}_{-0.0007}$ & $^{+0.0010}_{-0.0010}$ & $^{+0.0006}_{-0.0006}$ & $^{+0.0007}_{-0.0007}$ & $^{+0.0009}_{-0.0009}$ & $^{+0.0009}_{-0.0009}$ & $^{+0.0011}_{-0.0011}$ \TStrut\BStrut\\ \hline
$\Delta_{23}:$ $B_{\psi(2S)\rightarrow \mu^+\mu^-} \pm 1\sigma$ & $^{-0.0073}_{+0.0077}$ & $^{-0.0042}_{+0.0041}$ & $^{-0.0045}_{+0.0045}$ & $^{-0.0066}_{+0.0067}$ & $^{-0.0048}_{+0.0049}$ & $^{-0.0055}_{+0.0053}$ & $^{-0.0026}_{+0.0025}$ & $^{-0.0059}_{+0.0061}$ & $^{-0.0075}_{+0.0078}$ & $^{-0.0046}_{+0.0043}$ & $^{-0.0053}_{+0.0050}$ \TStrut\BStrut\\ \hline
$\Delta_{24}:$ $B_{\psi(2S)\rightarrow \mu^+\mu^-\pi^+\pi^-} \pm 1\sigma$ & $^{-0.0005}_{+0.0005}$ & $^{-0.0007}_{+0.0007}$ & $^{-0.0006}_{+0.0006}$ & $^{-0.0008}_{+0.0008}$ & $^{-0.0006}_{+0.0006}$ & $^{-0.0010}_{+0.0010}$ & $^{-0.0007}_{+0.0007}$ & $^{-0.0005}_{+0.0005}$ & $^{-0.0005}_{+0.0005}$ & $^{-0.0011}_{+0.0011}$ & $^{-0.0014}_{+0.0014}$ \TStrut\BStrut\\ \hline\hline
Total & $^{+0.0161}_{-0.0196}$ & $^{+0.0131}_{-0.0239}$ & $^{+0.0146}_{-0.0157}$ & $^{+0.0174}_{-0.0223}$ & $^{+0.0160}_{-0.0177}$ & $^{+0.0206}_{-0.0210}$ & $^{+0.0096}_{-0.0112}$ & $^{+0.0127}_{-0.0181}$ & $^{+0.0196}_{-0.0164}$ & $^{+0.0324}_{-0.0369}$ & $^{+0.0702}_{-0.0713}$ \TStrut\BStrut\\ \hline
\end{tabular}
\end{sideways}
\caption{Table of systematic uncertainties.  Uncertainties (see Section~\ref{sect:Systematics}) are shown for the individual variations for the final $R$ value as well as $R$ in 
bins of $W$ and $|t|$.  A series of dashes indicates that the variation led to an uncertainty in one direction only.}
\label{tab:syst}
\end{table}
%


\vspace{4cm}
\newpage
\pagebreak[4]

%
%
\end{document}